\documentclass[12pt,onecolumn]{article} \evensidemargin=0.20in
\usepackage{amsmath,amssymb,epsfig,psfrag}
\newcommand{\figref}[1]{Fig.~\ref{fig:#1}}
\newcommand{\Integer}{{\mathbb{Z}}}

\newcommand{\Complex}{{\mathbb{C}}}

\global\def\putFrag#1#2#3#4{
\begin{figure}[tp]
\begin{center}
#4 \epsfxsize=#3in \epsfbox{figures/#1.eps}
\end{center}
\caption{\small{#2}} \label{fig:#1}
\end{figure}
}

\newtheorem{theorem}{Theorem}

\newenvironment{proof}{{\sl Proof\/}:\ \ }

\begin{document}
\title{The Throughput-Reliability Tradeoff\\
in MIMO Channels\footnote {The authors are with the ECE Department
at the Ohio State University (email:azariany,
helgamal@ece.osu.edu).}}
\author{Kambiz~Azarian and Hesham~El~Gamal}

\maketitle

\begin{abstract}
\normalsize In this paper, an outage limited MIMO channel is
considered. We build on Zheng and Tse's elegant formulation of the
diversity-multiplexing tradeoff to develop a better understanding
of the asymptotic relationship between the probability of error,
transmission rate, and signal-to-noise ratio. In particular, we
identify the limitation imposed by the multiplexing gain notion
and develop a new formulation for the throughput-reliability
tradeoff that avoids this limitation. The new characterization is
then used to elucidate the asymptotic trends exhibited by the
outage probability curves of MIMO channels.
\end{abstract}
\section{Problem Formulation}
This paper revolves around the following question: What does a $3$
dB increase in the signal-to-noise ratio (SNR) buy in an outage
limited Multi-Input Multi-Output (MIMO) channel? In an Additive
White Gaussian Noise (AWGN) setting, it is well known that a $3$ dB
increase in SNR translates into an extra bit in channel capacity in
the high SNR regime. The scenario considered in this paper, however,
is more involved. We study an outage limited channel where the
randomness of the instantaneous mutual information results in a
non-zero lower bound on the probability of error, for non-zero
constant transmission rates. Hence, a fundamental tradeoff between
the throughput, as quantified by the transmission rate, and
reliability, as quantified by the so-called outage probability,
arises. Our work explores this tradeoff in the high SNR regime.

To be more specific, we consider a MIMO wireless communication
system with $m$ transmit and $n$ receive antennas. We adopt a
quasi-static flat-fading setup where the path gains remain constant
over $l$ consecutive symbol-intervals (\emph{i.e.,} a block), but
change independently from one block to another. We further assume a
coherent communication model implying the availability of channel
state information (CSI) only at the receiver. Under these
assumptions, the channel input-output relation is given by:
\begin{align}
\mathbf{y} &= \sqrt{\frac{\rho}{m}} \mathbf{H} \mathbf{x} +
\mathbf{w}. \label{eq:1}
\end{align}
In \eqref{eq:1}, $\mathbf{y} \in \Complex^n$ has entries $y_i$
representing the signal received at antenna $i \in
\{1,\cdots,n\}$, $\mathbf{x} \in \Complex^m$ has entries $x_j$
denoting the signal transmitted by antenna $j \in \{1,\cdots,m\}$,
$\mathbf{H} \in \Complex^{n \times m}$ has entries $h_{ij}$ which
represents the path gain between receive antenna
$i\in\{1,\cdots,n\}$ and transmit antenna $j\in\{1,\cdots,m\}$,
and $\mathbf{w} \in \Complex^n$ represents the unit-variance
additive white Gaussian noise. We model $\{h_{ij}\}$ as i.i.d
zero-mean and unit-variance complex Gaussian random variables.
Finally, $\rho$ corresponds to the SNR at each receive antenna.

Our work builds on Zheng and Tse's formulation of the
diversity-multiplexing tradeoff \cite{ZT02}. This formulation
assumes a family of space-time codes $\{{\cal C}_\rho\}$ indexed by
their operating SNR $\rho$, such that the code ${\cal C}_\rho$ has
rate $R(\rho)$, in bits per channel use (bpcu), and error
probability $P_e(\rho)$. For this family, the multiplexing gain $r$
and the diversity gain $d$ are defined by\footnote{Unless otherwise
stated, in this paper all logarithms are assumed to be in base $2$.}
\begin{align}
r \triangleq \lim_{\rho \to \infty} \frac{R(\rho)}{\log \rho}
\text{~~~~and~~~~} d \triangleq - \lim_{\rho \to \infty} \frac{\log
P_e(\rho)}{ \log \rho}. \label{multiplexing-rd}
\end{align}
The optimal diversity-multiplexing tradeoff yields the maximum
possible diversity gain for every value of $r$.  The main result
of \cite{ZT02} is summarized in the following theorem:

\begin{theorem} \label{zheng-tse-th1}
The optimal diversity gain for the coherent quasi-static MIMO
channel with $m$ transmit and $n$ receive antennas, at multiplexing
gain $r$, is given by $d(r) = f(r)$, where $f(\cdot)$ is the
piecewise linear function joining the points $(k,(m-k)(n-k))$ for $k
= 0,\ldots,\min\{m,n\}$. Moreover, there exists a code that achieves
$d(r)$ for all block lengths $l \geq m+n-1$.
\end{theorem}
In the sequel, we will use the notation $d_{max}=mn$ and
$r_{max}=\min\{m,n\}$. To motivate our work, we use the
diversity-multiplexing tradeoff to make a first attempt towards
answering our central question on the utility of a $3$ dB SNR gain
in quasi-static MIMO channels. Using the extreme points of the
tradeoff curve, i.e., $(0,d_{max})$ and $(r_{max},0)$, a {\bf
reasonable} conjecture is
\begin{enumerate}
\item At high enough SNRs, one can fix the transmission rate and
obtain $d_{max}$ orders of decay in the outage probability (on a
log-log scale) for every $10$ dB gain in SNR.

\item At high enough SNRs, one can fix the outage probability and
obtain a rate increase of $r_{max}$ bpcu for every $3$ dB gain in
SNR.
\end{enumerate}
\figref{outage_2x2_4,8} and \figref{outage_2x2_28,32} examine the
validity of this conjecture in a $2\times 2$ MIMO channel. In
these figures, the transmission rates and SNR ranges are carefully
chosen to illustrate the following points.
\begin{enumerate}
\item The slope of the outage probability curves in
\figref{outage_2x2_4,8} is shown to approach the asymptotic value
of $d_{max}=4$, on the log-log scale, as predicted by the first
part of our conjecture. The \emph{surprising} observation,
however, is that for a constant outage probability a $4.5$ dB gain
in SNR is needed to obtain an $r_{max}=2$ bpcu increase in the
throughput (To avoid fractions of a dB, the figure shows a $9$ dB
spacing for a $4$ bpcu throughput increase). This contrasts the
second part of our conjecture which predicts the need for only $3$
dB for every $2$ bpcu. More interestingly, this $4.5$ dB
\emph{horizontal spacing} seems to persist as the SNR increases.

\item \figref{outage_2x2_28,32}, on the other hand, comes
in close agreement with the second part of our conjecture. Here, the
horizontal spacing, for a $2$ bpcu increase in throughput, is seen
to be $3$ dB.  The disagreement in this case, however, is exhibited
in the fact that the slope of the outage probability curves,
corresponding to fixed rates, seems to \emph{stabilize} for a wide
range of SNRs at a value of $2$ (instead of $4$).

\item Repeating the experiment for different values of $m$ and $n$
reveals the same trends, i.e., 1) Our conjecture seems to offer
{\em partially} accurate predictions in certain {\em operating
regions}\footnote{A more formal definition of an operating region
is presented in the sequel.} and 2) Except for the $1 \times 1$
channel, the predictions for the outage probability rate of decay
and horizontal spacing are {\em never simultaneously} accurate.

\item Overall, these disagreements clearly disprove our
{\em naive} conjecture. However, it seems that the conjecture is
{\em not completely} false as it offers some accurate predictions,
at least in certain operating regions.
\end{enumerate}
Inspired by these observations, this paper aims at developing a
better understanding of the fundamental throughput-reliability
tradeoff in outage limited MIMO channels. It turns out that such an
understanding requires a more general formulation which is not
limited by the multiplexing gain notion as defined in
(\ref{multiplexing-rd}). In particular, the multiplexing gain notion
limits the scenarios of interest to {\em asymptotic lines} on the
$R-\log \rho$ plane as shown in \figref{lines}. Our formulation, on
the other hand, allows for investigating more general scenarios by
relaxing this constraint. Specifically, we shed more light on the
relationship between the three quantities $(R,\log \rho,
P_e(R,\rho)$), in the asymptotic limit of large $\rho$, when
\begin{align}
\limsup_{\rho \to \infty} \frac{R}{\log \rho} &\neq \liminf_{\rho
\to \infty} \frac{R}{\log \rho}. \label{no-mux}
\end{align}
It is clear that \eqref{no-mux} allows for investigating scenarios
defined by arbitrary asymptotic trajectories in the $R-\log \rho$
plane where the multiplexing gain is not defined
(\figref{trajectory} depicts such a trajectory). As argued in the
sequel, this {\em freedom} of walking along arbitrary trajectories,
on the $R-\log \rho$ plane, is the {\bf key} to obtaining accurate
predictions for the outage probability slopes and horizontal
spacings in different operating regions. While our characterization
is {\em rigorous} only in the asymptotic scenario where SNR grows to
infinity (i.e., $\rho \to \infty$), we will demonstrate, via
numerical results, that it yields very accurate predictions for
practically relevant values of SNR.

The rest of the paper is organized as follows. In
Section~\ref{trt}, we state our main result formulating the
throughput-reliability tradeoff (TRT) for the point-to-point
coherent MIMO channel, along with a sketch of the main ideas in
the proof. In this section, we also present numerical results and
intuitive arguments that demonstrate the utility of our results in
predicting the {\em behavior} of outage probability curves in the
high SNR regime. Section~\ref{apl} utilizes the TRT to shed more
light on the performance of various space-time architectures and
further extends our investigation to Automatic Repeat reQuest
(ARQ) channels. We offer few concluding remarks in
Section~\ref{conc}. In order to enhance the flow of the paper, the
proofs are collected in the Appendix.

\section{The Throughput-Reliability Tradeoff (TRT)}\label{trt}
An outage is defined as the event that the instantaneous mutual
information does not support the intended rate, i.e.,
\begin{align}
O_{p(\mathbf{x})} &\triangleq \{H \in \Complex^{n \times m}|
I(\mathbf{x};\mathbf{y}|\mathbf{H}=H) < R\}. \nonumber
\end{align}
Notice that the mutual information depends on both the channel
realization $H$ and the input distribution $p(\mathbf{x})$. The
outage probability $P_o(R,\rho)$ is then defined as
\begin{align}
P_o(R,\rho) &= \inf_{p(\mathbf{x})} \Pr \{ O_{p(\mathbf{x})} \}.
\nonumber
\end{align}
The following theorem characterizes the relationship between $R$,
$\rho$, and $P_o(R,\rho)$.

\begin{theorem} \label{thrm:1}
For the $m \times n$ MIMO channel described by \eqref{eq:1},
\begin{align}
\lim_{\begin{array}{c} \rho \to \infty \\
R \in \mathcal{R}(k) \end{array}} \frac{\log P_o(R,\rho) -
c(k)R}{\log \rho} &= -g(k), \label{eq:3}
\end{align}
where $P_o(R,\rho)$ denotes the outage probability at rate $R$ and
SNR $\rho$. $\mathcal{R}(k)$ is defined by
\begin{align}
\mathcal{R}(k) \triangleq \{R| k+1 > \frac{R}{\log \rho} > k \} &
\text{~~for~~} k \in \Integer, \min\{m,n\} > k \geq 0 \label{eq:4}.
\end{align}
In \eqref{eq:3}, $c(k)$ and $g(k)$ are given by
\begin{align}
c(k) &\triangleq m+n-(2k+1) \label{eq:5},
\end{align}
and
\begin{align}
g(k) &\triangleq mn-k(k+1) \label{eq:6}.
\end{align}
Moreover, in the degenerate case $R
> \min\{m,n\} \log \rho$, $ \lim_{\rho \to \infty} \log
P_o(R,\rho) / \log \rho =0$.
\end{theorem}
We refer to $g(k)$ as the {\em reliability gain coefficient} and
$t(k) \triangleq g(k)/c(k)$ as the {\em throughput gain
coefficient}.

\begin{proof}(Sketch)
Our proof  follows the same lines as the proof of
Theorem~\ref{zheng-tse-th1} in \cite{ZT02} except for the
fundamental challenge that the multiplexing gain is not defined
here. To handle this challenge, we judicially choose the region
${\cal R}(k)$ and find a lower bound on
\begin{align}
\liminf_{\begin{array}{c} \rho \to \infty \\
R \in \mathcal{R}(k) \end{array}} \frac{\log P_o(R,\rho) -
c(k)R}{\log \rho}, \nonumber
\end{align}
and an upper bound on
\begin{align}
\limsup_{\begin{array}{c} \rho \to \infty \\
R \in \mathcal{R}(k) \end{array}} \frac{\log P_o(R,\rho) -
c(k)R}{\log \rho} \nonumber
\end{align}
and show that the two bounds coincide for the choice of $c(k)$ and
$g(k)$ given by (\ref{eq:5}) and (\ref{eq:6}), respectively. The
detailed proof is reported in the Appendix.
\end{proof}

It is immediate to check that if the multiplexing gain is well
defined, i.e., if
\begin{align}
r = \lim_{\rho \to \infty} \frac{R(\rho)}{\log \rho} \nonumber
\end{align}
exists, then Theorem~\ref{thrm:1} reduces to
Theorem~\ref{zheng-tse-th1}. In the more general case, however,
Theorem~\ref{thrm:1} replaces the restrictive multiplexing gain
notion with the new concept of operating regions ${\cal R}(k)$. It
is worth noting that every operating region corresponds to a line
segment in the diversity-multiplexing tradeoff. In fact, this
correspondence inspires the following observation
\begin{align}
g(k)&=d(k)-kd^\prime(k^+),\label{eq:54} \\
c(k)&=-d^\prime(k^+), \label{eq:201}
\end{align}
where $d^\prime(k^+)$ is the slope of the line segment connecting
$d(k)$ and $d(k+1)$.

We are now ready to investigate the asymptotic trends of the
throughput-reliability tradeoff using Theorem~\ref{thrm:1}. The
following discussion hinges on the intuitive interpretation of
equation~(\ref{eq:3}).
\begin{align}
\log P_o(R,\rho) \approx c(k) R-g(k)\log \rho, \label{eq:203}
\end{align}
where the approximation in (\ref{eq:203}) becomes progressively
more accurate as $\rho$ increases. Equation (\ref{eq:203}) implies
that the slope of the outage curve, for a constant rate, is given
by $g(k)$, while the horizontal spacing in dB between two outage
curves with a $\Delta R$ rate difference is given by $3 \Delta R /
t(k)$. The key observation here is that in order to fix the
transmission rate (or outage probability), while staying in the
same operating region, one {\bf must} deviate from the linear
trajectory imposed by the multiplexing gain notion. The following
{\em heuristic} derivation of (\ref{eq:54}) and (\ref{eq:201})
further illustrates this point. To derive \eqref{eq:54}, we start
from two approximate relationships obtained from the
diversity-multiplexing tradeoff
\begin{align}
\log P_o\left(R,2^{\log \rho}\right) \approx -d\left(\frac{R}{\log
\rho}\right) \log \rho, \label{eq:23}
\end{align}
\begin{align}
\log P_o\left(R,2^{\log \rho + \Delta \log \rho}\right) &\approx
-d\left(\frac{R}{\log \rho + \Delta \log \rho}\right) (\log \rho +
\Delta \log \rho). \label{eq:32}
\end{align}
We further approximate $d\left(\frac{R}{\log \rho + \Delta \log
\rho}\right)$ with the first two terms of its Taylor series
expansion, i.e.,
\begin{align}
d\left(\frac{R}{\log \rho + \Delta \log \rho}\right) &\approx
d\left(\frac{R}{\log \rho}\right) - \frac{R \times \Delta \log
\rho}{\log \rho (\log \rho + \Delta \log \rho)}
d^{\prime}\left(\frac{R}{\log \rho}\right), \label{eq:53}
\end{align}
Now \eqref{eq:53}, together with \eqref{eq:23} and \eqref{eq:32},
gives
\begin{align}
\frac{\log P_o(R,2^{\log \rho}) - \log P_o(R,2^{\log \rho + \Delta
\log \rho})}{\Delta \log \rho} &\approx d\left(\frac{R}{\log
\rho}\right) - \frac{R}{\log \rho} d^\prime\left(\frac{R}{\log
\rho}\right). \label{eq:55}
\end{align}
Realizing that the left-hand side of \eqref{eq:55} gives the slope
of $P_o(R,\rho)$ with respect to $\rho$, i.e. $g(k)$, and that
$d(r)-rd^{\prime}(r)$ remains constant over the line segments of
$d(r)$, we get \eqref{eq:54}. Deriving \eqref{eq:201} follows the
same lines. In particular, we first compute the horizontal spacing,
$\Delta \log \rho$, between the outage curves corresponding to rates
$R$ and $R + \Delta R$. For this purpose, we use \eqref{eq:23} to
write
\begin{align}
\log P_o\left(R + \Delta R,2^{\log \rho + \Delta \log \rho}\right)
&\approx -d\left(\frac{R + \Delta R}{\log \rho + \Delta \log
\rho}\right) (\log \rho + \Delta \log \rho) \label{eq:57}
\end{align}
Then we expand $d(\frac{R + \Delta R}{\log \rho + \Delta \log
\rho})$ in a way similar to \eqref{eq:53} and equate \eqref{eq:23}
with \eqref{eq:57} to get
\begin{align}
\Delta \log \rho &\approx \frac{-d^{\prime} (\frac{R}{\log \rho})
}{d(\frac{R}{\log \rho}) - \frac{R}{\log \rho}
d^\prime(\frac{R}{\log \rho})} \Delta R, \nonumber \\
\Delta \log \rho &\approx \frac{-d^{\prime} (\frac{R}{\log
\rho})}{g(k)} \Delta R. \label{eq:56}
\end{align}
Realizing that $\Delta \log \rho= \frac{c(k)}{g(k)} \Delta R$, and
that $d^{\prime}(r)$ remains constant over the line segments of
$d(r)$, we get \eqref{eq:201}.

Revisiting our naive conjecture, we can now see that
$g(0)=mn=d_{max}$ which agrees with the first part, while
$t(\min\{n,m\}-1)=\min\{m,n\}=r_{max}$ agrees with the second part.
This explains the {\em partial} correctness of the conjecture and
the fact that, except for the $1 \times 1$ MIMO channel, the two
parts are {\bf never} simultaneously accurate, since they correspond
to different operating regions. We also observe that both the
reliability and throughput gain coefficients exhibit a staircase
behavior. Moreover, it is easy to see that $g(k)$ is a decreasing
function of $k$, while $t(k)$ is an increasing function of $k$. This
implies that, at a fixed rate $R$ and for sufficiently large SNRs,
as $R / \log \rho$ increases (i.e., $\rho$ decreases), the decay
rate of $P_o(R,\rho)$ decreases and the horizontal spacing between
the outage curves corresponding to a fixed rate difference shrinks.
In the following, we present numerical results that validate this
observation.

Before proceeding to the numerical results, we need the following
rule of thump for determining the operating regions, for large but
finite values of $\rho$ and $R$, such that the approximation in
(\ref{eq:203}) is accurate. The operating point $(R,\rho)$ is in
operating region ${\cal R}(k)$ if and only if
\begin{align}
\frac{\rho^k}{2^R} \leq \delta, \text{~~and~~}
\frac{2^R}{\rho^{k+1}}\leq \delta, \nonumber
\end{align}
where $\delta$ is a \textbf{small} value which determines the
accuracy of the approximation. It is now straightforward to see
that the high SNR segment of \figref{outage_2x2_4,8} falls in the
region ${\cal R}(0)$ and, indeed, the $4$ levels of diversity and
$4.5$ dB spacing (for every $2$ bpcu throughput increase) in this
figure correspond precisely to $g(0)=4$ and $t(0)=4/3$. Similarly,
the high SNR segment of \figref{outage_2x2_28,32} falls in ${\cal
R}(1)$ and, again, the $2$ levels of diversity and $3$ dB spacing,
for every $2$ bpcu throughput increase, agree with $g(1)=2$ and
$t(1)=2$. \figref{outage_2x2_20} and \figref{outage_2x2_20,24} are
different from the previous two figures in that the high SNR
segments of the outage curves fall within \textbf{both} of the two
regions. As a result, the slope of the curves and the spacing
between them change as the operating point leaves one operating
region and enters another. Again, the values of the slopes,
spacings, and operating points at which the change occurs (which
can be read from \figref{regions_2x2_20}) match nicely with the
predictions of the TRT as formulated by Theorem \ref{thrm:1}.
\figref{outage_3x3_4,10} through \figref{outage_3x3_34,40}
correspond to a $3 \times 3$ MIMO system. As can be seen from
\figref{outage_3x3_4,10}, the solid segment of the curve
corresponding to $R=10$ bpcu, falls in $\mathcal{R}(1)$ and, as
predicted, we observe a slope of $g(1)=7$. It should be noted,
however, that the tail of the curve corresponding to $R=4$ bpcu is
leaving $\mathcal{R}(1)$ and entering $\mathcal{R}(0)$ and thus
the slope of this curve is larger than $7$ (about 7.7). For the
same reason, the horizontal spacing between the two curves (almost
$10$ dB) is larger than the value predicted for $k=1$ (i.e., $7.7$
dB). The solid segments of the outage curves corresponding to
$R=58$ and $64$ bpcu in \figref{outage_3x3_58,64} fall in
$\mathcal{R}(2)$, and therefore, we observe $3$ levels of
diversity and $3$ dB of spacing, for every $3$ bpcu throughput
increase, which correspond precisely to $g(2)=3$ and $t(2)=3$.
\figref{outage_3x3_40} and \figref{outage_3x3_34,40} depict the
case where the high SNR segments fall within two different regions
($k=2$ and $k=1$). Again the slopes and spacings are in agreement
with the predictions of the TRT.

\section{Applications}\label{apl}
The following result establishes the operational significance of
Theorem~\ref{thrm:1} by showing that the optimal space-time code
probability of error exhibits the same asymptotic behavior as the
outage probability.

\begin{theorem} \label{thrm:2}
The probability of error for the {\bf optimal} coding/decoding
scheme used in conjunction with channel (\ref{eq:1}) satisfies
\begin{align}
\lim_{\begin{array}{c} \rho \to \infty \\
R \in \mathcal{R}(k) \end{array}} \frac{\log P_e(R,\rho) -
c(k)R}{\log \rho} &= -g(k), \label{eq:206}
\end{align}

where ${\cal R}(k)$, $c(k)$, and $g(k)$ are given by (\ref{eq:4}),
(\ref{eq:5}), and (\ref{eq:6}), respectively. Moreover, there
exists a coding scheme that achieves (\ref{eq:206}) for $l \geq
m+n-1$.
\end{theorem}
\begin{proof} (Sketch) The proof follows the same lines as \cite{ZT02}. In
particular, the converse is obtained via a careful use of Fano's
inequality. The achievability is established using an ensemble of
Gaussian codebooks along with the appropriate use of the union
bound.
\end{proof}

One can also derive the TRT achievable by certain suboptimal
space-time architectures. In this paper, we restrict our study to
square V-BLAST protocols and orthogonal space-time constellations.
In the V-BLAST architecture, the input stream is split into $m$
sub-streams. These sub-streams are then encoded independently and
transmitted over the $m$ transmit antennas \cite{FGVW:99}. The
following theorem characterizes the throughput-reliability
tradeoff for this protocol when a maximum likelihood decoder is
employed.

\begin{theorem} \label{thrm:4}
The ML error probability for a V-BLAST communication system with
$m$ transmit and $m$ receive antennas satisfies\footnote{The
subscript ``vb'' stands for V-BLAST.}
\begin{align}
\lim_{\begin{array}{c} \rho \to \infty \\
R \in \mathcal{R}_{vb} \end{array}} \frac{\log P_e(R,\rho) - R}{\log
\rho} &= -m, \label{eq:59}
\end{align}
where $\mathcal{R}_{vb}$ is given by
\begin{align}
\mathcal{R}_{vb} &\triangleq  \{R| m
> \frac{R}{\log \rho} > 0 \} \nonumber.
\end{align}
Moreover, there exists a coding scheme that achieves \eqref{eq:59}
for $l \geq 2m-1$.
\end{theorem}
\begin{proof}
Please refer to the appendix for a detailed proof.
\end{proof}

\figref{outage_vblast_2x2_ml_8,12} depicts the outage curves
corresponding to $R=8$ and $12$ bpcu for a $2 \times 2$ V-BLAST
scheme with ML decoding. As can be seen from this figure, the high
SNR segments of the outage curves achieve $2$ levels of diversity
with a horizontal spacing of $6$ dB. These values agree with
$g_{vb}(0)=2$ and $t_{vb}(0)=2$.
\figref{outage_mimo_vblast_4,16,32}, on the other hand, compares the
outage behavior of $2 \times 2$ MIMO and ML V-BLAST schemes for
$R=4,16$ and $32$ bpcu. As can be seen from this figure, the outage
curves for the two schemes coincide while $ R / \log \rho
> 1$. In particular, the curves corresponding to $R=32$ bpcu are
almost identical. However, for $R/ \log \rho <1$, the sub-optimality
of the V-BLAST becomes evident. In fact, the curve corresponding to
the $2 \times 2$ MIMO with $R=4$ bpcu approaches $4$ levels of
diversity very rapidly, while the curve corresponding to the V-BLAST
only attains $2$ levels.

Similarly, orthogonal space-time constellations allow for a simple
TRT characterization. An orthogonal constellation of size $m$,
length $l$, and rate $k/l$ (in symbols per channel use (spcu)) is
a space-time code $\mathbf{X} \in \Complex^{m \times l}$ such that
\begin{align}
\mathbf{X} \mathbf{X}^H &= \left( \sum_{i=1}^k |x_i|^2 \right)
\times \mathbf{I}_m, \label{eq:75}
\end{align}
where $\{x_i\}_{i=1}^k$ denote the symbols to be sent,
$\mathbf{I}_m$ is the $m \times m$ identity matrix and
$\mathbf{X}^H$ denotes the hermitian of matrix $\mathbf{X}$
\cite{TJC:99}. As an example, consider the orthogonal
constellation with $m=2$, $l=2$, and rate one, which is known as
the Alamouti code \cite{A:98}. In this case
\begin{align}
\mathbf{X} &= \begin{bmatrix} x_1 & -x_2^* \\ x_2 & x_1^*
\end{bmatrix}, \nonumber
\end{align}
where $x^*$ denotes the complex conjugate of $x$. Notice that
\begin{align}
\mathbf{X} \mathbf{X}^H &= (|x_1|^2+|x_2|^2) \times \mathbf{I}_2,
\nonumber
\end{align}
as required by \eqref{eq:75}. The received signal matrix
$\mathbf{Y} \in \Complex^{n \times l}$ at the destination can be
written as
\begin{align}
\mathbf{Y} &= \sqrt{\frac{\rho}{m}} \mathbf{H} \mathbf{X} +
\mathbf{W}. \label{eq:79}
\end{align}
The ML receiver performs linear processing on $\mathbf{Y}$ to
yield the following equivalent parallel channel model
\begin{align}
\tilde{y}_i &= \sqrt{\frac{\rho}{m} \| \mathbf{H} \|^2} x_i +
\tilde{w}_i, \text{~~for~~} i=1, \cdots, k. \label{eq:74}
\end{align}
In \eqref{eq:74}, $\| \mathbf{H} \|^2$ denotes the Frobenius norm
of $\mathbf{H}$ (i.e., $\| \mathbf{H} \|^2 \triangleq \sum
|h_{ij}|^2$, where $\{ h_{ij} \}$ are the entries of $\mathbf{H}$)
and $\{ \tilde{w}_i \}_{i=1}^k$ are i.i.d complex Gaussian random
variables of zero mean and unit variance. The following theorem
gives the throughput-reliability tradeoff for orthogonal
constellations.

\begin{theorem} \label{thrm:5}
The optimal throughput-reliability tradeoff for an orthogonal
constellation of size $m$, length $l$, rate $k/l$ spcu ($R$ bpcu)
and $n$ receive antennas satisfies\footnote{The subscript ``oc''
stands for an orthogonal constellation.}
\begin{align}
\lim_{\begin{array}{c} \rho \to \infty \\
R \in \mathcal{R}_{oc} \end{array}} \frac{\log P_e(R,\rho)
-\frac{l}{k}mnR}{\log \rho} &= -mn, \label{eq:76}
\end{align}
where $\mathcal{R}_{oc}$ is given by
\begin{align}
\mathcal{R}_{oc} &\triangleq \{R| \frac{k}{l}
> \frac{R}{\log \rho} > 0 \} \nonumber,
\end{align}
Moreover, there exists an outer coding scheme (one that maps the
information bits into symbols $\{x_i\}_{i=1}^k$) that achieves
\eqref{eq:76} for $k \geq mn$.
\end{theorem}
\begin{proof}
The proof follows immediately from Theorem \ref{thrm:2} and the
fact that the orthogonal constellation of interest effectively
converts the underlying $m \times n$ MIMO channel of rate $R$ (as
given by \eqref{eq:79}) into a $1 \times mn$ channel of rate
$\frac{l}{k}R$ (as given by \eqref{eq:74}).
\end{proof}

\figref{outage_alamouti_2x2_4,8} depicts the outage curves
corresponding to $R=4$ and $8$ bpcu for a $2 \times 2$ Alamouti
scheme. As can be seen from this figure, the high SNR segments of
the outage curves achieve $4$ levels of diversity with a
horizontal spacing of $12$ dB. These values agree with
$g_{oc}(0)=4$ and $t_{oc}(0)=1$.
\figref{outage_mimo_alamouti_4,16,32}, on the other hand, compares
the outage behavior of the $2 \times 2$ MIMO channel and the
Alamouti constellation for $R=4,16$ and $32$ bpcu. As can be seen
from this figure, while the outage curves for the two schemes
coincide for small rates, the sub-optimality of the Alamouti
scheme becomes evident at higher values of $R$. In particular, the
curves corresponding to $R=4$ bpcu are almost identical and for
$R=32$ bpcu the curve corresponding to the Alamouti scheme lags
that of MIMO by more than $40$ dB.

Finally, we extend our analysis to MIMO-ARQ channels. In this setup,
the transmitter starts by picking up a message from the transmission
buffer. It then uses a space-time encoder to map the message to a
sequence of blocks $\mathbf{X}_p \in \Complex^{m \times l}, L \geq p
\geq 1$. During transmission-round $p$, the transmitter sends
$\mathbf{X}_p$ one column at a time over its $m$ antennas. The
receiver then tries to decode the message. If successful, it sends
back a positive acknowledgement signal (ACK), which causes the
transmitter to start sending the next message. However, if the
receiver detects an error, it requests another round of transmission
by feeding back a negative acknowledgement signal (NACK). The only
exception to this rule is when $L$ rounds of transmission have
already been sent, in which case the transmitter abandons sending
the current message and goes to the next one. In this paper, we
address the long-term static channel scenario, where all of the
transmission-rounds corresponding to a message take place over the
same channel realization. We also impose a short term power
constraint on the transmitter, such that power-control is not
possible \cite{ECD:04}. At this point, we need to distinguish
between two closely related parameters, namely, the first-round
transmission rate and the long-term average throughput. Assume that
each message consists of $b$ information bits which means that the
first-round transmission rate is $R_1=b/l$ bpcu. Since some of the
messages take more than one transmission-round to be sent, the
long-term average throughput $\eta$ is strictly less than $R_1$. The
gap between the two quantities, however, diminishes as the SNR
grows. This is due to the fact that at high SNRs, most of the
messages are decoded error-free after the first round of
transmission and the ARQ retransmission-rounds are used only for
those \emph{rare} events in which the message does not get through
with only one round of transmission. Recognizing the operational
significance of $\eta$, in the following we state the TRT for ARQ
channels in terms of $\eta$, rather than $R_1$.

\begin{theorem} \label{thrm:3}
The optimal throughput-reliability tradeoff for the coherent
block-fading MIMO ARQ channel with $m$ transmit antennas, $n$
receive antennas, $L$ maximum number of transmission-rounds, under
the long-term static channel and short-term power constraint
assumptions is given by\footnote{The subscript ``ls'' stands for
long-term static.}
\begin{align}
\lim_{\begin{array}{c} \rho \to \infty \\
\eta \in \mathcal{R}_{ls}(k) \end{array}} \frac{\log P_e(\eta,\rho)
- c_{ls}(k)\eta}{\log \rho} &= -g_{ls}(k), \label{eq:58}
\end{align}
where $\eta$ denotes the long-term average data rate. In
\eqref{eq:58}, $\mathcal{R}_{ls}(k)$, $c_{ls}(k)$ and $g_{ls}(k)$
are defined by
\begin{align}
\mathcal{R}_{ls}(k) &\triangleq \left\{ \begin{array}{lll} \{\eta|
(k+1)L > \frac{\eta}{\log \rho} > kL \} & \text{for} & k \in
\Integer, \lfloor
\frac{\min\{m,n\}}{L} \rfloor > k \geq 0 \\
\{\eta| \min\{m,n\} > \frac{\eta}{\log \rho} > \lfloor
\frac{\min\{m,n\}}{L} \rfloor L \} & \text{for} & k = \lfloor
\frac{\min\{m,n\}}{L} \rfloor
\end{array} \nonumber \right.,
\end{align}
\begin{align}
c_{ls}(k) \triangleq \frac{c(k)}{L} \text{~~and~~} g_{ls}(k)
&\triangleq g(k) \nonumber,
\end{align}
respectively. $c(k)$ and $g(k)$ are given by \eqref{eq:5} and
\eqref{eq:6}.
\end{theorem}

\begin{proof} (Sketch) The proof follows the same lines as that of
Theorem $5$ in \cite{ECD:04}. In particular, the converse is
obtained by lower-bounding the error probability of the ARQ protocol
with that of a ML decoder that operates on the whole codeword $\{
\mathbf{X_p} \}_{p=1}^L$. The achievability, on the other hand, is
established through the use of an ensemble of Gaussian code-books,
along with a bounded-distance decoder. The main idea here is to
differentiate between the undetected-errors (i.e., the ones for
which the receiver sends back an ACK signal) and those errors that
the decoder makes after requesting $L$ rounds of transmission. It
can then be shown that, through judicious choice of decoder
threshold-distance, the latter error type becomes dominant, and
hence, the lower and upper bounds become tight. It is then
straightforward to argue the existence of codes in the ensemble that
perform at least as well as the ensemble average.
\end{proof}

\section{Conclusions}\label{conc}
We have developed a new asymptotic relationship between $P_e$,
$\rho$, and $R$ in outage limited MIMO channels. By relaxing the
restriction imposed by the multiplexing gain notion, our
characterization sheds more light on the throughput-reliability
tradeoff in the high SNR regime. We presented numerical results
which validate our claim that, the throughput-reliability tradeoff
offers accurate predictions on the {\em worth} of a $3$ dB SNR
gain in a MIMO wireless system operating in the high SNR regime.
For our results to be valid, the only requirement is that the
operating point be within certain well defined regions in the
$R-\log\rho$ plane. Characterizing the performance in the
transitional regions remains an open problem.

\section{Acknowledgment}
The authors would like to thank Profs.~G.~Caire and M.~O.~Damen
for inspiring discussions.

\section{Appendix} \label{app}

\subsection{Proof of Theorem~\ref{thrm:1}}
For the channel described by \eqref{eq:1}, an outage is defined as
the event that the realized mutual information does not support the
intended rate, i.e.
\begin{align}
O_{p(\mathbf{x})} &\triangleq \{H \in \Complex^{n \times m}|
I(\mathbf{x};\mathbf{y}|\mathbf{H}=H) < R\}. \label{eq:40}
\end{align}
Notice that the mutual information depends on both, the realized
channel $H$ and the input distribution $p(\mathbf{x})$. For this
channel, the outage probability $P_o(R,\rho)$ is defined as
\begin{align}
P_o(R,\rho) &= \inf_{p(\mathbf{x})} \Pr \{ O_{p(\mathbf{x})} \}.
\nonumber
\end{align}
It is shown in \cite{ZT02} that
\begin{align}
P_o(R,\rho) &\leq \Pr\{\log \det(I_n + \frac{\rho}{m} HH^H) < R\} \text{~~and} \nonumber \\
P_o(R,\rho) &\ge \Pr\{\log \det(I_n + \rho HH^H) < R\}. \nonumber
\end{align}
These equations can be re-written as
\begin{align}
P_o(R,\rho) &\leq \Pr\{\log (\prod_{i=1}^{\min\{m,n\}}
(1+\frac{\rho}{m} \mu_i)) < R\} \text{~~and} \label{eq:7}\\
P_o(R,\rho) &\ge \Pr\{\log (\prod_{i=1}^{\min\{m,n\}} (1+\rho
\mu_i)) < R\}, \label{eq:2}
\end{align}
where $\mu_{\min\{m,n\}} \geq \cdots \geq \mu_1 \geq 0$ represent
the ordered eigenvalues of $HH^H$. The joint Probability Density
Function (PDF) of $(\mu_1,\cdots,\mu_{\min\{m,n\}})$ is given by the
Wishart distribution, i.e.
\begin{align}
p(\mathbf{\mu}) &= K_{m,n}^{-1} \prod_{i=1}^{\min\{m,n\}}
\mu_i^{|m-n|} \prod_{i<j} (\mu_i-\mu_j)^2 e^{-\sum_{i} \mu_i},
\label{eq:8}
\end{align}
where $\mathbf{\mu} \triangleq (\mu_1,\cdots,\mu_{\min\{m,n\}})$ and
$K_{m,n}$ is a normalizing factor. Now, let us focus on \eqref{eq:7}
and introduce the change of variables
\begin{align}
\alpha_i &\triangleq \frac{\log (1+ \frac{\rho}{m} \mu_i)}{R}.
\label{eq:10}
\end{align}
This implies that $\alpha_{\min\{m,n\}} \geq \cdots \geq \alpha_1
\geq 0$. In terms of the new variables, \eqref{eq:7} can be written
as
\begin{align}
P_o(R,\rho) &\leq \Pr\{ \ \mathcal{A} \}, \label{eq:11}
\end{align}
where
\begin{align}
\mathcal{A} &\triangleq \{ \mathbf{\alpha}| \alpha_{\min\{m,n\}}
\geq \cdots \geq \alpha_1 \geq 0, 1 - \sum_i \alpha_i > 0 \}.
\label{eq:9}
\end{align}
In \eqref{eq:9}, $\mathbf{\alpha} \triangleq (\alpha_1, \cdots,
\alpha_{\min\{m,n\}})$. On the other hand, \eqref{eq:8} becomes
\begin{align}
p(\mathbf{\alpha}) =& K R^{\min\{m,n\}} \rho^{-mn} 2^{R \sum_i
\alpha_i} \times \nonumber \\
&\prod_{i=1}^{\min\{m,n\}} (2^{\alpha_i R}-1)^{|m-n|} \prod_{i<j}
(2^{\alpha_i R}-2^{\alpha_j R})^2 e^{-\sum_i \frac{m(2^{\alpha_i
R}-1)}{\rho}}, \label{eq:12}
\end{align}
where $K \triangleq K_{m,n}^{-1} (\ln 2)^{\min\{m,n\}} m^{mn}$.
Next, we define $\mathcal{R}_{\delta}(k)$, for integer $k$'s, as
\begin{align}
\mathcal{R}_{\delta}(k) &\triangleq \left\{ \begin{array}{lll} \{R|
\frac{1}{\delta} > \frac{\log \rho}{R} > 1 + \delta \} & \text{if} & k = 0 \\
\{R| \frac{1}{k}-\delta > \frac{\log \rho}{R} > \frac{1}{k+1} +
\delta \} & \text{if} & k \in \Integer, \min\{m,n\} > k > 0
\end{array} \label{eq:17} \right.,
\end{align}
where $\delta$ denotes a small positive value. Notice that
$\delta=0$ reduces $\mathcal{R}_{\delta}(k)$ to $\mathcal{R}(k)$, as
given by \eqref{eq:4}. Now, it follows from \eqref{eq:11} that
\begin{align}
P_o(R,\rho) 2^{-c(k)R} &\leq 2^{-c(k)R} \int_{\mathcal{A}}
p(\mathbf{\alpha}) \text{d} \mathbf{\alpha},\text{~~} R \in
\mathcal{R}_{\delta}(k) \nonumber
\end{align}
Note that this expression is true, regardless of the choice for
$c(k)$, i.e., at this point we regard $c(k)$ as an \emph{arbitrary}
function of $k$. This inequality can be written as
\begin{align}
P_o(R,\rho) 2^{-c(k)R} &\leq A_1(R,\rho,\epsilon) +
A_2(R,\rho,\epsilon),\text{~~} R \in \mathcal{R}_{\delta}(k)
\nonumber
\end{align}
where
\begin{align}
A_1(R,\rho,\epsilon) &\triangleq 2^{-c(k)R} \int_{\mathcal{A}_1}
p(\mathbf{\alpha}) \text{d} \mathbf{\alpha}, &A_2(R,\rho,\epsilon)
\triangleq 2^{-c(k)R} \int_{\mathcal{A}_2} p(\mathbf{\alpha})
\text{d} \mathbf{\alpha} \nonumber
\end{align}
and
\begin{align}
\mathcal{A}_1 &\triangleq \{\mathbf{\alpha} \in \mathcal{A} |
\alpha_{\min\{m,n\}} > \frac{\log \rho}{R}+\epsilon \},
&\mathcal{A}_2 \triangleq \{\mathbf{\alpha} \in \mathcal{A}|
\frac{\log \rho}{R}+\epsilon \geq \alpha_{\min\{m,n\}} \}.
\label{eq:14}
\end{align}
This means that
\begin{align}
\limsup_{\begin{array}{c} \rho \to \infty \\
R \in \mathcal{R}_{\delta}(k) \end{array}} \frac{\log P_o(R,\rho)
-c(k)R}{\log \rho} \leq& \limsup_{\begin{array}{c} \rho \to \infty \\
R \in \mathcal{R}_{\delta}(k) \end{array}} \frac{ \log \big(1+
A_1(R,\rho,\epsilon) / A_2(R,\rho,\epsilon) \big)}{\log \rho} +
\nonumber \\
&\limsup_{\begin{array}{c} \rho \to \infty \\
R \in \mathcal{R}_{\delta}(k) \end{array}} \frac{\log
A_2(R,\rho,\epsilon)}{\log \rho}, \label{eq:13}
\end{align}
To characterize the first term in the right-hand side of
\eqref{eq:13}, we notice that
\begin{align}
A_1(R,\rho,\epsilon) \leq& KR^{\min\{m,n\}} e^{\frac{m}{\rho}}
\rho^{-mn} 2^{-c(k)R} \times \nonumber \\
&\int_{\mathcal{A}_1} 2^{R \sum_i \alpha_i}
\prod_{i=1}^{\min\{m,n\}} 2^{\alpha_i |m-n| R} \prod_{i<j} 2^{2
\alpha_j R} e^{-m2^{(\alpha_{\min\{m,n\}} - \frac{\log \rho}{R})R}}
\text{d} \mathbf{\alpha} \nonumber \\
\leq& KR^{\min\{m,n\}} e^{\frac{m}{\rho}} \rho^{-mn} 2^{-c(k)R}
\int_{\mathcal{A}_1} 2^{f(\mathbf{\alpha})R} e^{-m2^{\epsilon R}}
\text{d} \mathbf{\alpha}, \nonumber
\end{align}
where
\begin{align}
f(\mathbf{\alpha}) &\triangleq \sum_{i=1}^{\min\{m,n\}} (|m-n|+2i-1)
\alpha_i. \label{eq:30}
\end{align}
Realizing that $\text{Vol}\{\mathcal{A}_1\} \leq 1$, we conclude
\begin{align}
A_1(R,\rho,\epsilon) \leq& KR^{\min\{m,n\}} (2^{\epsilon
R})^{\frac{f_1 - c(k)}{\epsilon}} e^{-m2^{\epsilon R}}
e^{\frac{m}{\rho}} \rho^{-mn}, \label{eq:18}
\end{align}
where
\begin{align}
f_1 &\triangleq \sup_{\mathcal{A}_1} f(\mathbf{\alpha}). \nonumber
\end{align}
On the other hand
\begin{align}
A_2(R,\rho,\epsilon) \ge& K R^{\min\{m,n\}}
e^{\frac{m\min\{m,n\}}{\rho}} \rho^{-mn} 2^{-c(k)R}
\int_{\mathcal{A}_2} e^{-\sum_i m 2^{-(\frac{\log \rho}{R}
-\alpha_i)R}} 2^{R
\sum_i \alpha_i} \times \nonumber \\
&\prod_{i=1}^{\min\{m,n\}} (1-2^{-\alpha_i R})^{|m-n|}
2^{|m-n|\alpha_i R} \prod_{i<j} (1-2^{-(\alpha_j-\alpha_i)R})^2
2^{2\alpha_j R} \text{d} \mathbf{\alpha}, \nonumber \\
\geq& K R^{\min\{m,n\}} e^{\frac{m\min\{m,n\}}{\rho}} \rho^{-mn}
2^{-c(k)R} \int_{\mathcal{A}_{\epsilon_1}} e^{-m\min\{m,n\} 2^{-
\epsilon_1 R}} 2^{R \sum_i \alpha_i} \times \nonumber \\
&\prod_{i=1}^{\min\{m,n\}} (1-2^{-\epsilon_1 R})^{|m-n|}
2^{|m-n|\alpha_i R} \prod_{i<j} (1-2^{-\epsilon_1 R})^2 2^{2\alpha_j
R} \text{d} \mathbf{\alpha}, \label{eq:80}
\end{align}
where
\begin{align}
\mathcal{A}_{\epsilon_1} &\triangleq \{ \mathbf{\alpha} \in
\mathcal{A}_2| \frac{\log \rho}{R} - \alpha_{ \min\{m,n\}} >
\epsilon_1, \alpha_1
> \epsilon_1, |\alpha_j - \alpha_i|
> \epsilon_1 ~\forall i \neq j \}. \nonumber
\end{align}
Realizing that $e^{-2^{-\epsilon_1 R}} \geq (1-2^{-\epsilon_1 R})$,
\eqref{eq:80} yields
\begin{align}
A_2(R,\rho,\epsilon) &\ge K R^{\min\{m,n\}}
e^{\frac{m\min\{m,n\}}{\rho}} (1-2^{-\epsilon_1
R})^{m(n+\min\{m,n\})} \rho^{-mn} 2^{-c(k)R}
\int_{\mathcal{A}_{\epsilon_1}} 2^{f(\mathbf{\alpha})R} \text{d}
\mathbf{\alpha}, \nonumber
\end{align}
where, as before, $f(.)$ is given by \eqref{eq:30}. Let us define
$\mathbf{\alpha}^*$ as
\begin{align}
\mathbf{\alpha}^* &\triangleq \arg \sup_{ \mathbf{\alpha} \in
\mathcal{A}_{\epsilon_1} } f(\mathbf{\alpha}). \nonumber
\end{align}
Then it follows from the continuity of $f(.)$ that, for any
$\epsilon_2
> 0$, there exists a neighborhood $I_{\epsilon_2}$ of
$\mathbf{\alpha}^*$, within which
\begin{align}
f(\mathbf{\alpha}) &\ge f(\mathbf{\alpha}^*) - \epsilon_2. \nonumber
\end{align}
This means that
\begin{align}
A_2(R,\rho,\epsilon) \ge& K R^{\min\{m,n\}}
e^{\frac{m\min\{m,n\}}{\rho}} (1-2^{-\epsilon_1
R})^{m(n+\min\{m,n\})} \rho^{-mn} 2^{-c(k)R}
\int_{\mathcal{A}_{\epsilon_1} \cap I_{\epsilon_2}}
2^{f(\mathbf{\alpha})R} \text{d} \mathbf{\alpha}, \nonumber \\
A_2(R,\rho,\epsilon) \ge& K R^{\min\{m,n\}}
e^{\frac{m\min\{m,n\}}{\rho}} (1-2^{-\epsilon_1
R})^{m(n+\min\{m,n\})} \rho^{-mn}
2^{(f(\mathbf{\alpha}^*) -c(k) - \epsilon_2)R} \times \nonumber \\
&\text{Vol}\{\mathcal{A}_{\epsilon_1} \cap I_{\epsilon_2}\}.
\label{eq:81}
\end{align}
Now, from \eqref{eq:18}, \eqref{eq:81} and the fact that
$e^{-\frac{m(\min\{m,n\}-1)}{\rho}} \leq 1$, we conclude that
\begin{align}
\frac{A_1(R,\rho,\epsilon)}{A_2(R,\rho,\epsilon)} \leq&
\big(1-(2^{\epsilon R})^{-\frac{\epsilon_1}{\epsilon}}
\big)^{-m(n+\min\{m,n\})} (2^{\epsilon R})^{\frac{f_1 -
f(\mathbf{\alpha}^*) + \epsilon_2}{\epsilon}} e^{-m(2^{\epsilon R})}
\text{Vol}^{-1} \{\mathcal{A}_{\epsilon_1} \cap I_{\epsilon_2} \}.
\label{eq:82}
\end{align}
This means that
\begin{align}
\limsup_{\begin{array}{c} \rho \to \infty \\
R \in \mathcal{R}_{\delta}(k) \end{array}} \frac{ \log \big(1+
A_1(R,\rho,\epsilon) / A_2(R,\rho,\epsilon) \big)}{\log \rho} = 0
\label{eq:19}
\end{align}
Note that \eqref{eq:19} holds, whether $\rho$ growing to infinity
and $R \in \mathcal{R}_{\delta}(k)$ result in $R$ growing to
infinity or not. This is because the right hand side of
\eqref{eq:82} decays \emph{exponentially} with $2^{\epsilon R}$,
while it only grows \emph{polynomially} with the same variable. To
characterize the second term on the right-hand side of
\eqref{eq:13}, we note that
\begin{align}
A_2(R,\rho,\epsilon) \leq& K R^{\min\{m,n\}} \rho^{-mn} 2^{-c(k)R}
\int_{\mathcal{A}_2} 2^{R \sum_i \alpha_i} \prod_{i=1}^{\min\{m,n\}}
2^{\alpha_i |m-n| R} \prod_{i<j} 2^{2 \alpha_j
R} \text{d} \mathbf{\alpha}, \nonumber \\
\leq& K R^{\min\{m,n\}} 2^{(f_2 - c(k))R} \rho^{-mn}. \nonumber
\end{align}
This means that
\begin{align}
\frac{\log A_2(R,\rho,\epsilon)}{\log \rho } &\leq \frac{\log (K
R^{\min\{m,n\}})}{\log \rho} + (f_2 - c(k)) \frac{R}{\log \rho} -mn,
\label{eq:20}
\end{align}
where
\begin{align}
f_2 &\triangleq \sup_{\mathcal{A}_2} f(\mathbf{\alpha}). \nonumber
\end{align}
To derive $f_2$, one needs to consider two different cases. The
first case is when $R \in \mathcal{R}_{\delta}(0)$, in which case
\begin{align}
f_2 &= m+n-1,\text{~~} R \in \mathcal{R}_{\delta}(0). \label{eq:15}
\end{align}
In this case, the supremum is achieved at
$\mathbf{\alpha}^*=(0,\cdots,0,1)$. The second case is when $R \in
\mathcal{R}_{\delta}(k)$, for $k \in \Integer$ and $\min\{m,n\} > k
>0$, where
\begin{align}
f_2 &= [m+n-(2k+1)]+k(k+1) (\frac{\log \rho}{R}+\epsilon), &R \in
\mathcal{R}_{\delta}(k), \min\{m,n\} > k >0. \label{eq:16}
\end{align}
The supremum happens at
\begin{align}
\mathbf{\alpha}^* &= (0,\cdots,1-k (\frac{\log \rho}{R}+\epsilon),
\underbrace{\frac{\log \rho}{R}+\epsilon ,\cdots, \frac{\log
\rho}{R}+\epsilon}_{k \text{~~times}}). \nonumber
\end{align}
Notice that, assuming $\epsilon \leq \delta$, \eqref{eq:17}
guarantees that $1-k (\frac{\log \rho}{R}+\epsilon) > 0$. Plugging
in for $f_2$ in \eqref{eq:20}, we conclude
\begin{align}
\frac{\log A_2(R,\rho,\epsilon)}{\log \rho} \leq& \frac{\log (K
R^{\min\{m,n\}})}{\log \rho} - g(k) + (mn-g(k))
\frac{R}{\log \rho} \epsilon + \nonumber \\
&\left(\tilde{c}(k)-c(k)\right) \frac{R}{\log \rho}, \text{~~for~~}
R \in \mathcal{R}_{\delta}(k). \label{eq:22}
\end{align}
In this expression, $g(k)$ is given by \eqref{eq:6} and
$\tilde{c}(k)$ is defined as
\begin{align}
\tilde{c}(k) &\triangleq m+n-(2k+1) \label{eq:24}.
\end{align}
Now, from \eqref{eq:13}, \eqref{eq:19} and \eqref{eq:22}, together
with the fact that $\epsilon$ can be made arbitrarily small, one
concludes
\begin{align}
\limsup_{\begin{array}{c} \rho \to \infty \\
R \in \mathcal{R}_{\delta}(k) \end{array}} \frac{\log P_o(R,\rho)
- c(k)R}{\log \rho} \leq -g(k)& + \nonumber \\
\left(\tilde{c}(k) - c(k)\right) \times \limsup_{\begin{array}{c}
\rho \to \infty \\ R \in \mathcal{R}_{\delta}(k)
\end{array}} \frac{R}{\log \rho}&. \label{eq:25}
\end{align}
Next we turn our attention to \eqref{eq:2} and introduce the
following change of variables.
\begin{align}
\beta_i &\triangleq \frac{\log (1+ \rho \mu_i)}{R}.
\end{align}
This implies that $\beta_{\min\{m,n\}} \geq \cdots \geq \beta_1 \geq
0$. In terms of the new variables, \eqref{eq:2} can be written as
\begin{align}
P_o(R,\rho) &\ge \Pr\{ \mathcal{B} \}, \label{eq:26}
\end{align}
where
\begin{align}
\mathcal{B} &\triangleq \{ \mathbf{\beta}| \beta_{\min\{m,n\}} \geq
\cdots \geq \beta_1 \geq 0, 1 - \sum_i \beta_i > 0 \}. \label{eq:27}
\end{align}
In \eqref{eq:27}, $\mathbf{\beta} \triangleq (\beta_1, \cdots,
\beta_{\min\{m,n\}})$. On the other hand, \eqref{eq:8} becomes
\begin{align}
p(\mathbf{\beta}) =& K R^{\min\{m,n\}} \rho^{-mn} 2^{R \sum_i
\beta_i} \times \nonumber \\
&\prod_{i=1}^{\min\{m,n\}} (2^{\beta_i R}-1)^{|m-n|} \prod_{i<j}
(2^{\beta_i R}-2^{\beta_j R})^2 e^{-\sum_i \frac{2^{\beta_i
R}-1}{\rho}}, \label{eq:28}
\end{align}
where $K \triangleq K_{m,n}^{-1} (\ln 2)^{\min\{m,n\}}$. This means
that
\begin{align}
P_o(R,\rho) 2^{-c(k)R} &\ge 2^{-c(k)R} \int_{\mathcal{B}}
p(\mathbf{\beta}) \text{d} \mathbf{\beta},\text{~~} R \in
\mathcal{R}_{\delta}(k). \nonumber
\end{align}
Thus
\begin{align}
P_o(R,\rho) 2^{-c(k)R} \ge& K R^{\min\{m,n\}}
e^{\frac{\min\{m,n\}}{\rho}} \rho^{-mn} 2^{-c(k)R}
\int_{\mathcal{B}} e^{-\sum_i 2^{-(\frac{\log \rho}{R} -\beta_i)R}}
2^{R
\sum_i \beta_i} \times \nonumber \\
&\prod_{i=1}^{\min\{m,n\}} (1-2^{-\beta_i R})^{|m-n|}
2^{|m-n|\beta_i R} \prod_{i<j} (1-2^{-(\beta_j-\beta_i)R})^2
2^{2\beta_j R} \text{d} \mathbf{\beta}, \nonumber \\
\geq& K R^{\min\{m,n\}} e^{\frac{\min\{m,n\}}{\rho}} \rho^{-mn}
2^{-c(k)R} \int_{\mathcal{B}_{\epsilon_1}} e^{-\min\{m,n\} 2^{-
\epsilon_1 R}} 2^{R \sum_i \beta_i} \times \nonumber \\
&\prod_{i=1}^{\min\{m,n\}} (1-2^{-\epsilon_1 R})^{|m-n|}
2^{|m-n|\beta_i R} \prod_{i<j} (1-2^{-\epsilon_1 R})^2 2^{2\beta_j
R} \text{d} \mathbf{\beta}, \label{eq:29}
\end{align}
where
\begin{align}
\mathcal{B}_{\epsilon_1} &\triangleq \{ \mathbf{\beta} \in
\mathcal{B}| \frac{\log \rho}{R} - \epsilon_1 > \beta_{
\min\{m,n\}}, \beta_1 > \epsilon_1, |\beta_j - \beta_i|
> \epsilon_1 ~\forall i \neq j \}. \nonumber
\end{align}
Realizing that $e^{-2^{-\epsilon_1 R}} \geq (1-2^{-\epsilon_1 R})$,
\eqref{eq:29} yields
\begin{align}
P_o(R,\rho) 2^{-c(k)R} &\ge K R^{\min\{m,n\}}
e^{\frac{\min\{m,n\}}{\rho}} (1-2^{-\epsilon_1 R})^{mn+\min\{m,n\}}
\rho^{-mn} 2^{-c(k)R} \int_{\mathcal{B}_{\epsilon_1}}
2^{f(\mathbf{\beta})R} \text{d} \mathbf{\beta}, \nonumber
\end{align}
where, as before, $f(.)$ is given by \eqref{eq:30}. Let us define
$\mathbf{\beta}^*$ as
\begin{align}
\mathbf{\beta}^* &\triangleq \arg \sup_{ \mathbf{\beta} \in
\mathcal{B}_{\epsilon_1} } f(\mathbf{\beta}). \nonumber
\end{align}
Again, it follows from the continuity of $f(.)$ that, for any
$\epsilon_2
> 0$, there exists a neighborhood $I_{\epsilon_2}$ of
$\mathbf{\beta}^*$, within which
\begin{align}
f(\mathbf{\beta}) &\ge f(\mathbf{\beta}^*) - \epsilon_2. \nonumber
\end{align}
This means that
\begin{align}
P_o(R,\rho) 2^{-c(k)} \ge& K R^{\min\{m,n\}}
e^{\frac{\min\{m,n\}}{\rho}} (1-2^{-\epsilon_1 R})^{mn+\min\{m,n\}}
\rho^{-mn} 2^{-c(k)R} \int_{\mathcal{B}_{\epsilon_1} \cap
I_{\epsilon_2}}
2^{f(\mathbf{\beta})R} \text{d} \mathbf{\beta}, \nonumber \\
P_o(R,\rho) 2^{-c(k)} \ge& K R^{\min\{m,n\}}
e^{\frac{\min\{m,n\}}{\rho}} (1-2^{-\epsilon_1 R})^{mn+\min\{m,n\}}
\rho^{-mn} 2^{(f(\mathbf{\beta}^*) -c(k) - \epsilon_2)R} \times
\nonumber \\ &\text{Vol}\{\mathcal{B}_{\epsilon_1} \cap
I_{\epsilon_2}\}. \nonumber
\end{align}
Thus
\begin{align}
\liminf_{\begin{array}{c} \rho \to \infty \\
R \in \mathcal{R}_{\delta}(k) \end{array}} \frac{\log P_o(R,\rho)
- c(k)R}{\log \rho} \ge -mn& + \nonumber \\
\liminf_{\begin{array}{c} \rho \to \infty \\
R \in \mathcal{R}_{\delta}(k) \end{array}} (f(\mathbf{\beta}^*)
-c(k) - \epsilon_2) \frac{R}{\log \rho}&. \label{eq:31}
\end{align}
Since \eqref{eq:31} is valid for arbitrarily small values of
$\epsilon_1$ and $\epsilon_2$, we conclude
\begin{align}
\liminf_{\begin{array}{c} \rho \to \infty \\
R \in \mathcal{R}_{\delta}(k) \end{array}} \frac{\log P_o(R,\rho)
- c(k)R}{\log \rho} \ge -g(k)& + \nonumber \\
\left( \tilde{c}(k) -c(k) \right) \times \liminf_{\begin{array}{c}
\rho \to \infty \\ R \in \mathcal{R}_{\delta}(k)
\end{array}} \frac{R}{\log \rho}&, \label{eq:33}
\end{align}
where we have used the fact that $\mathcal{B}_0 =
\mathcal{A}_2|_{\epsilon=0}$, which means that $f(\mathbf{\beta}^*)$
can be easily derived from \eqref{eq:15} and \eqref{eq:16} by simply
plugging in $\epsilon=0$. Examining \eqref{eq:25} and \eqref{eq:33}
reveals that the choice
\begin{align}
c(k)= \tilde{c}(k), \text{~~} \forall k \label{eq:34}
\end{align}
guarantees the existence of
\begin{align}
\lim_{\begin{array}{c} \rho \to \infty \\
R \in \mathcal{R}_{\delta}(k) \end{array}} \frac{\log P_o(R,\rho) -
c(k)R}{\log \rho} &= -g(k), \label{eq:47}
\end{align}
\emph{regardless} of whether $\lim_{\rho \to \infty} \frac{R}{\log
\rho}$ exits or not. Now, since \eqref{eq:47} holds for arbitrarily
small values of $\delta$, we get \eqref{eq:3}. Note that
\eqref{eq:34}, together with \eqref{eq:24}, result in \eqref{eq:5}
and thus complete the proof.

\subsection{Proof of Theorem \ref{thrm:2}}
The proof follows that of Theorem $2$ in \cite{ZT02}. In particular,
we prove \eqref{eq:206} in two steps. The first step is to show that
\begin{align}
\liminf_{\begin{array}{c} \rho \to \infty \\
R \in \mathcal{R}(k) \end{array}} \frac{\log P_e(R,\rho) -
c(k)R}{\log \rho} &\ge -g(k). \label{eq:41}
\end{align}
Towards this end, let us denote the ML error probability,
conditioned on a certain channel realization $H$, by $P_{E|H}$. It
then follows that
\begin{align}
P_e(R,\rho) &= \text{E}_H\{P_{E|H}(R,\rho)\}, \nonumber \\
&= \int P_{E|H}(R,\rho) p(H) \text{d}H. \nonumber
\end{align}
Thus
\begin{align}
P_e(R,\rho) &\geq \int_\mathcal{C} P_{E|H}(R,\rho) p(H) \text{d}H,
\label{eq:35}
\end{align}
where $\mathcal{C}$ denotes any subset of the set of all channel
realizations. Let us define $\mathcal{C}_{\epsilon}$ as the set of
channel realizations for which the conditional ML error probability
cannot be made smaller than $\epsilon$, i.e.
\begin{align}
\mathcal{C}_{\epsilon} &\triangleq \{H | P_{E|H}(R,\rho) \geq
\epsilon \}. \label{eq:36}
\end{align}
From \eqref{eq:35} and \eqref{eq:36} one concludes that
\begin{align}
P_e(R,\rho) &\geq \epsilon P_{\mathcal{C}_{\epsilon}}(R,\rho),
\nonumber
\end{align}
where
\begin{align}
P_{\mathcal{C}_{\epsilon}}(R,\rho) &=
\text{Pr}\{\mathcal{C}_{\epsilon}\}. \label{eq:37}
\end{align}
This means that
\begin{align}
\frac{\log P_e(R,\rho) - c(k)R}{\log \rho} &\geq \frac{\log
P_{\mathcal{C}_{\epsilon}}(R,\rho) - c(k)R}{\log \rho} + \frac{\log
\epsilon}{\log \rho}, \nonumber
\end{align}
where $c(k)$ is given by \eqref{eq:5}, or
\begin{align}
\liminf_{\begin{array}{c} \rho \to \infty \\
R \in \mathcal{R}_{\delta}(k) \end{array}} \frac{\log P_e(R,\rho)
- c(k)R}{\log \rho} &\geq \liminf_{\begin{array}{c} \rho \to \infty \\
R \in \mathcal{R}_{\delta}(k) \end{array}} \frac{\log
P_{\mathcal{C}_{\epsilon}}(R,\rho) - c(k)R}{\log \rho},
\label{eq:38}
\end{align}
where $\mathcal{R}_{\delta}(k)$ is given by \eqref{eq:17}. Now,
application of Fano's inequality reveals that
\begin{align}
P_{E|H}(R,\rho) \geq 1 - \frac{\text{I}(\mathbf{x};\mathbf{y}
|\mathbf{H}=H)}{R} - \frac{1}{Rl}, \nonumber
\end{align}
where $l$ denotes the codeword length \cite{ZT02}. This, together
with \eqref{eq:36}, means that
\begin{align}
\{H| 1 - \frac{\text{I}(\mathbf{x};\mathbf{y} | \mathbf{H}=H)}{R} -
\frac{1}{Rl} \geq \epsilon \} &\subseteq \mathcal{C}_{\epsilon},
\nonumber
\end{align}
which using \eqref{eq:40} results in
\begin{align}
P_{\mathcal{C}_{\epsilon}}(R,\rho) &\geq P_o((1 - \epsilon -
\frac{1}{Rl})R,\rho) \text{~~or~~} \nonumber \\
\liminf_{\begin{array}{c} \rho \to \infty \\
R \in \mathcal{R}_{\delta}(k) \end{array}} \frac{\log
P_{\mathcal{C}_{\epsilon}}(R,\rho) - c(k)R}{\log \rho} &\geq
\liminf_{\begin{array}{c} \rho \to \infty \\
R \in \mathcal{R}_{\delta}(k) \end{array}} \frac{\log
P_o((1-\epsilon)R,\rho) - c(k)R}{\log \rho}. \label{eq:39}
\end{align}
Now, from \eqref{eq:38} and \eqref{eq:39}, together with the fact
that both, $\delta$ and $\epsilon$ can be made arbitrarily small, we
conclude that
\begin{align}
\liminf_{\begin{array}{c} \rho \to \infty \\
R \in \mathcal{R}(k) \end{array}} \frac{\log P_e(R,\rho)
- c(k)R}{\log \rho} &\geq \liminf_{\begin{array}{c} \rho \to \infty \\
R \in \mathcal{R}(k) \end{array}} \frac{\log P_o(R,\rho) -
c(k)R}{\log \rho}. \nonumber
\end{align}
But, from Theorem \ref{thrm:1} (refer to \eqref{eq:3}), we know that
the right-hand side equals $-g(k)$ (given by \eqref{eq:6}). This
proves \eqref{eq:41} and thus completes the first step.

The second step in proving \eqref{eq:206} is to show that
\begin{align}
\limsup_{\begin{array}{c} \rho \to \infty \\
R \in \mathcal{R}(k) \end{array}} \frac{\log P_e(R,\rho) -
c(k)R}{\log \rho} &\leq -g(k), \label{eq:42}
\end{align}
provided that the codeword length, $l$, satisfies $l \geq m+n-1$. To
prove this, consider a Gaussian code-book with $2^{Rl}$ codewords of
length $l$. It is straightforward to verify that the ML error
probability, conditioned on a certain channel realization, is
upper-bounded by
\begin{align}
P_{E|H}(R,\rho) &\leq 2^{Rl} \det(I_n + \frac{\rho}{2m} HH^H)^{-l},
\nonumber \\
&= 2^{Rl} \prod_{i=1}^{\min\{m,n\}} (1 + \frac{\rho}{2m}
\mu_i)^{-l}, \label{eq:44}
\end{align}
where $\mu_{\min\{m,n\}} \geq \cdots \geq \mu_1 \geq 0$ represent
the ordered eigenvalues of $HH^H$. The joint PDF of
$(\mu_1,\cdots,\mu_{\min\{m,n\}})$ is given by \eqref{eq:8}. The
change of variables
\begin{align}
\gamma_i &\triangleq \frac{\log(1+\frac{\rho}{2m}\mu_i)}{R},
\nonumber
\end{align}
changes \eqref{eq:44} and \eqref{eq:8}, into
\begin{align}
P_{E|\mathbf{\gamma}}(R,\rho) &\leq 2^{(1 - \sum_{i=1}^{\min\{m,n\}}
\gamma_i )Rl} \label{eq:45}
\end{align}
and
\begin{align}
p(\mathbf{\gamma}) =& K R^{\min\{m,n\}} \rho^{-mn} 2^{R \sum_i
\gamma_i} \times \nonumber \\
&\prod_{i=1}^{\min\{m,n\}} (2^{\gamma_i R}-1)^{|m-n|} \prod_{i<j}
(2^{\gamma_i R}-2^{\gamma_j R})^2 e^{-\sum_i \frac{2m(2^{\gamma_i
R}-1)}{\rho}}, \nonumber
\end{align}
where $\gamma \triangleq (\gamma_1, \cdots, \gamma_{\min\{m,n\}})$
and $K \triangleq K_{m,n}^{-1} (\ln2)^{\min\{m,n\}} (2m)^{mn}$. Next
we define $\mathcal{D}$ as
\begin{align}
\mathcal{D} &\triangleq \{ \mathbf{\gamma}| \gamma_{\min\{m,n\}}
\geq \cdots \geq \gamma_i \geq 0, 1 - \sum_{i=1}^{\min\{m,n\}}
\gamma_i \geq 0 \}. \nonumber
\end{align}
Referring to \eqref{eq:45} reveals that $\mathcal{D}$ consists of
those channel realizations for which the upper-bound on the ML error
probability cannot be made arbitrarily small, even through the use
of infinitely long codewords. For these channel realizations, we
upper-bound $P_{E|\mathbf{\gamma}}(R,\rho)$ by $1$, i.e.
\begin{align}
P_e(R,\rho) &= P_{E,\mathcal{D}^c}(R,\rho) +
P_{E,\mathcal{D}}(R,\rho), \nonumber \\
P_e(R,\rho) &\leq P_{E,\mathcal{D}^c}(R,\rho) +
P_{\mathcal{D}}(R,\rho), \label{eq:43}
\end{align}
where $\mathcal{D}^c$ denotes the complement of $\mathcal{D}$. Let
us first focus on $P_{\mathcal{D}}(R,\rho)$. Realizing that
$\mathcal{D}$ is precisely the same set as $\mathcal{A}$ (refer to
\eqref{eq:9}) and that, up to a scaling factor, $p(\mathbf{\gamma})$
is identical to $p(\mathbf{\alpha})$ (refer to \eqref{eq:12}), it
follows immediately that
\begin{align}
\limsup_{\begin{array}{c} \rho \to \infty \\
R \in \mathcal{R}(k) \end{array}} \frac{\log P_{\mathcal{D}}(R,\rho)
- c(k)R}{\log \rho} &\leq -g(k). \label{eq:48}
\end{align}
Now, turning our attention back to $P_{E,\mathcal{D}^c}(R,\rho)$, we
realize that
\begin{align}
\lim_{\begin{array}{c} \rho \to \infty \\
R \in \mathcal{R}(k) \end{array}} P_{E,\mathcal{D}^c_1}(R,\rho)
2^{-c(k)R} &= 0, \label{eq:49}
\end{align}
which means that
\begin{align}
\limsup_{\begin{array}{c} \rho \to \infty \\
R \in \mathcal{R}(k) \end{array}} \frac{\log
P_{E,\mathcal{D}^c}(R,\rho) - c(k)R}{\log \rho} &=
\limsup_{\begin{array}{c} \rho \to \infty \\
R \in \mathcal{R}(k) \end{array}} \frac{\log
P_{E,\mathcal{D}^c_2}(R,\rho) - c(k)R}{\log \rho}, \label{eq:50}
\end{align}
where
\begin{align}
\mathcal{D}^c_1 \triangleq \{ \mathbf{\gamma} \notin \mathcal{D}|
\gamma_{\min\{m,n\}} > \frac{\log \rho}{R} \} \text{~~and~~}
\mathcal{D}^c_2 \triangleq \{ \mathbf{\gamma} \notin \mathcal{D}|
\gamma_{\min\{m,n\}} \leq \frac{\log \rho}{R} \}. \nonumber
\end{align}
Notice that \eqref{eq:49} holds for exactly the same reason as
\eqref{eq:19} does (refer to the comment after \eqref{eq:19}). Now,
using \eqref{eq:45}, we have
\begin{align}
P_{E,\mathcal{D}^c_2}(R,\rho) 2^{-c(k)R} &= 2^{-c(k)R}
\int_{\mathcal{D}^c_2} P_{E|\mathbf{\gamma}}(R,\rho)
p(\mathbf{\gamma}) \text{d}
\mathbf{\gamma}, \nonumber \\
&\leq K R^{\min\{m,n\}} \rho^{-mn} 2^{-c(k)R} \int_{\mathcal{D}^c_2}
2^{[f(\mathbf{\gamma})+l(1 - \sum_{i=1}^{\min\{m,n\}} \gamma_i )]R}
\text{d} \mathbf{\gamma}, \nonumber
\end{align}
where
\begin{align}
f(\mathbf{\gamma}) &\triangleq 2^{\sum_{i} (|m-n|+2i-1) \gamma_i R}.
\nonumber
\end{align}
Thus
\begin{align}
P_{E,\mathcal{D}^c_2}(R,\rho) 2^{-c(k)R} &\leq K R^{\min\{m,n\}}
\rho^{-mn} 2^{(f_2-c(k))R} \text{Vol}\{\mathcal{D}^c_2\},
\label{eq:51}
\end{align}
where
\begin{align}
f_2 &\triangleq \sup_{\mathcal{D}^c_2} f(\mathbf{\gamma})+l(1 -
\sum_{i=1}^{\min\{m,n\}} \gamma_i ). \nonumber
\end{align}
Realizing that for $l \geq m+n-1$, the supremum occurs at
$\mathbf{\gamma}=\mathbf{\gamma}^*$, such that $1-\sum \gamma^*_i
=0$, $f_2$ can be easily derived from \eqref{eq:15} and
\eqref{eq:16} by simply plugging in $\epsilon=0$. Therefore,
\eqref{eq:51} gives
\begin{align}
\limsup_{\begin{array}{c} \rho \to \infty \\
R \in \mathcal{R}(k) \end{array}} \frac{\log
P_{E,\mathcal{D}^c_2}(R,\rho) - c(k)R}{\log \rho} &\leq -g(k).
\label{eq:52}
\end{align}
Now, from \eqref{eq:52}, \eqref{eq:50} and \eqref{eq:48}, we
conclude \eqref{eq:42}, which together with \eqref{eq:41}, proves
\eqref{eq:206}. Since we proved \eqref{eq:206} using an
\emph{ensemble} of Gaussian codes, it follows that, for any
code-length $l \ge m+n-1$, there exists at least a code, for which
\eqref{eq:206} holds. This completes the proof.

\subsection{Proof of Theorem \ref{thrm:4}}
Realizing that the V-BLAST protocol essentially transforms the $m
\times m$ MIMO channel into a multiple-access channel with $m$
single-antenna users and a destination with $m$ receive antennas, we
prove \eqref{eq:59} by following the same lines as that of Theorem
$2$ in \cite{TVZ:03}. In particular, let $E_S$ denote the event that
a certain decoder makes errors in decoding the codewords transmitted
by a subset $S$ of the antennas. It then follows that
\begin{align}
\sum_{S \neq \varnothing} \text{Pr}\{E_S\} \geq P_e(R,\rho).
\label{eq:65}
\end{align}
It is also clear that
\begin{align}
P_e(R,\rho) \geq \text{Pr}\{E_{S^*}\}, \label{eq:68}
\end{align}
where $S^*$ denotes any non-empty subset of $\{1, \cdots, m\}$. Let
us define $S^*$ as the non-empty subset of $\{1, \cdots, m\}$, such
that for all other non-empty subsets $S$, we have
\begin{align}
\liminf_{\begin{array}{cc} \rho \to \infty \\ \frac{|S|}{m}R \in
\mathcal{R}_S(k)\\ \frac{|S^*|}{m}R \in \mathcal{R}_{S^*}(k^*)
\end{array}} \Big( g_S(k) - g_{S^*}(k^*) \Big) - \Big(
\frac{|S|}{m}c_S(k) - \frac{|S^*|}{m}c_{S^*}(k^*) \Big)
\frac{R}{\log \rho} \geq 0. \label{eq:66}
\end{align}
In \eqref{eq:66}, $|S|$ denotes the cardinality of set $S$. Also,
$\mathcal{R}_S(k)$, $c_{S}(k)$ and $g_{S}(k)$ denote
$\mathcal{R}(k)$, $c(k)$ and $g(k)$, as defined by \eqref{eq:4},
\eqref{eq:5} and \eqref{eq:6}, for a MIMO channel with $|S|$
transmit and $m$ receive antennas. The proof of \eqref{eq:59} then
follows in three steps. First, we prove that for any decoder
\begin{align}
\liminf_{\begin{array}{cc} \rho \to \infty \\ \frac{|S^*|}{m}R \in
\mathcal{R}_{S^*}(k^*) \end{array}} \frac{P_e(R,\rho) -
\frac{|S^*|}{m}c_{S^*}(k) R }{\log \rho} \geq -g_{S^*}(k^*).
\label{eq:67}
\end{align}
This follows immediately from \eqref{eq:68} and the fact that
$\text{Pr}\{E_{S^*}\}$ upper-bounds the ML error probability of a
MIMO channel with $|S^*|$ transmit antennas, $m$ receive antennas
and rate $\frac{|S^*|}{m}R$. The second step in proving
\eqref{eq:59} is to show that there exists a code, along with a
decoder, for which
\begin{align}
\limsup_{\begin{array}{cc} \rho \to \infty \\ \frac{|S^*|}{m}R \in
\mathcal{R}_{S^*}(k^*) \end{array}} \frac{P_e(R,\rho) -
\frac{|S^*|}{m} c_{S^*}(k^*) R }{\log \rho} \leq -g_{S^*}(k^*).
\label{eq:69}
\end{align}
We prove this by showing that the error probability of the joint ML
decoder, averaged over the ensemble of Gaussian codes, satisfies
\eqref{eq:69}. The existence of the desired code then follows from
the fact that there exist codes in the ensemble that perform at
least as well as the average. For this purpose, assume that each of
the antennas uses a Gaussian code-book of code-word length $l=2m+1$
and size $2^{\frac{R}{m}l}$ codewords. It then follows from Theorem
\ref{thrm:2}, (refer to \eqref{eq:206}), that
\begin{align}
\lim_{\begin{array}{c} \rho \to \infty \\
\frac{|S|}{m}R \in \mathcal{R}_S(k) \end{array}} \frac{\log
\text{Pr} \{E_S\} - \frac{|S|}{m} c_S(k) R}{\log \rho} &= -g_S(k),
\text{~~} \forall S \neq \varnothing, \label{eq:72}
\end{align}
which means
\begin{align}
\lim_{\begin{array}{cc} \rho \to \infty \\ \frac{|S|}{m}R \in
\mathcal{R}_S(k)\\ \frac{|S^*|}{m}R \in \mathcal{R}_{S^*}(k^*)
\end{array}} \Bigg[ \frac{\log \text{Pr}\{E_S\} / \text{Pr}\{E_{S^*}\} }{\log
\rho} + \Big( g_S(k) - g_{S^*}(k^*) \Big)& - \nonumber \\
\Big( \frac{|S|}{m}c_S(k) - \frac{|S^*|}{m}c_{S^*}(k^*) \Big)&
\frac{R}{\log \rho} \Bigg]= 0. \label{eq:70}
\end{align}
Now, \eqref{eq:70}, together with \eqref{eq:66}, results in
\begin{align}
\limsup_{\begin{array}{cc} \rho \to \infty \\ \frac{|S|}{m}R \in
\mathcal{R}_S(k)\\ \frac{|S^*|}{m}R \in \mathcal{R}_{S^*}(k^*)
\end{array}}  \frac{\log \text{Pr}\{E_S\} / \text{Pr}\{E_{S^*}\} }{\log
\rho} \leq 0, \text{~~} \forall S \neq \varnothing. \label{eq:71}
\end{align}
Returning to \eqref{eq:65}, we have
\begin{align}
\frac{\log (1 + \sum_{S \neq S^*, \varnothing} \frac{
\text{Pr}\{E_S\}}{\text{Pr}\{E_{S^*}\}})}{\log \rho} + \frac{\log
\text{Pr}\{E_{S^*}\} - \frac{|S^*|}{m} c_{S^*}(k^*) R }{\log \rho}
&\geq \frac{\log P_e(R,\rho) - \frac{|S^*|}{m} c_{S^*}(k^*) R }{\log
\rho}. \nonumber
\end{align}
Taking the $\limsup$ of both sides, together with \eqref{eq:71} and
\eqref{eq:72} results in \eqref{eq:69}. Notice that \eqref{eq:67}
and \eqref{eq:69} mean that
\begin{align}
\lim_{\begin{array}{cc} \rho \to \infty \\ \frac{|S^*|}{m}R \in
\mathcal{R}_{S^*}(k^*) \end{array}} \frac{P_e(R,\rho) -
\frac{|S^*|}{m}c_{S^*}(k) R }{\log \rho} = -g_{S^*}(k^*).
\label{eq:73}
\end{align}
The third and last step in proving \eqref{eq:59} is to show that
$|S^*|=1$. One can prove this directly using the definitions of
$\mathcal{R}(k)$, $c(k)$ and $g(k)$ (equations \eqref{eq:4},
\eqref{eq:5} and \eqref{eq:6}). However, we choose to do this using
observations \eqref{eq:54} and \eqref{eq:201}. In particular, notice
that based on these observations, \eqref{eq:66} reduces to finding
the subset $S^*$, for which
\begin{align}
d_S(\frac{|S|}{m}r)-d_{S^*}(\frac{|S^*|}{m}r) \geq 0, \text{~~}
\forall S \neq \varnothing, \nonumber
\end{align}
where $d_S(r)$ represents the diversity-multiplexing tradeoff for a
MIMO system with $|S|$ transmit and $m$ receive antennas. From the
proof of Theorem $3$ in \cite{TVZ:03}, we know that $|S^*|=1$. Now,
this together with \eqref{eq:73} results in \eqref{eq:59} and thus
completes the proof.

\newpage
\putFrag{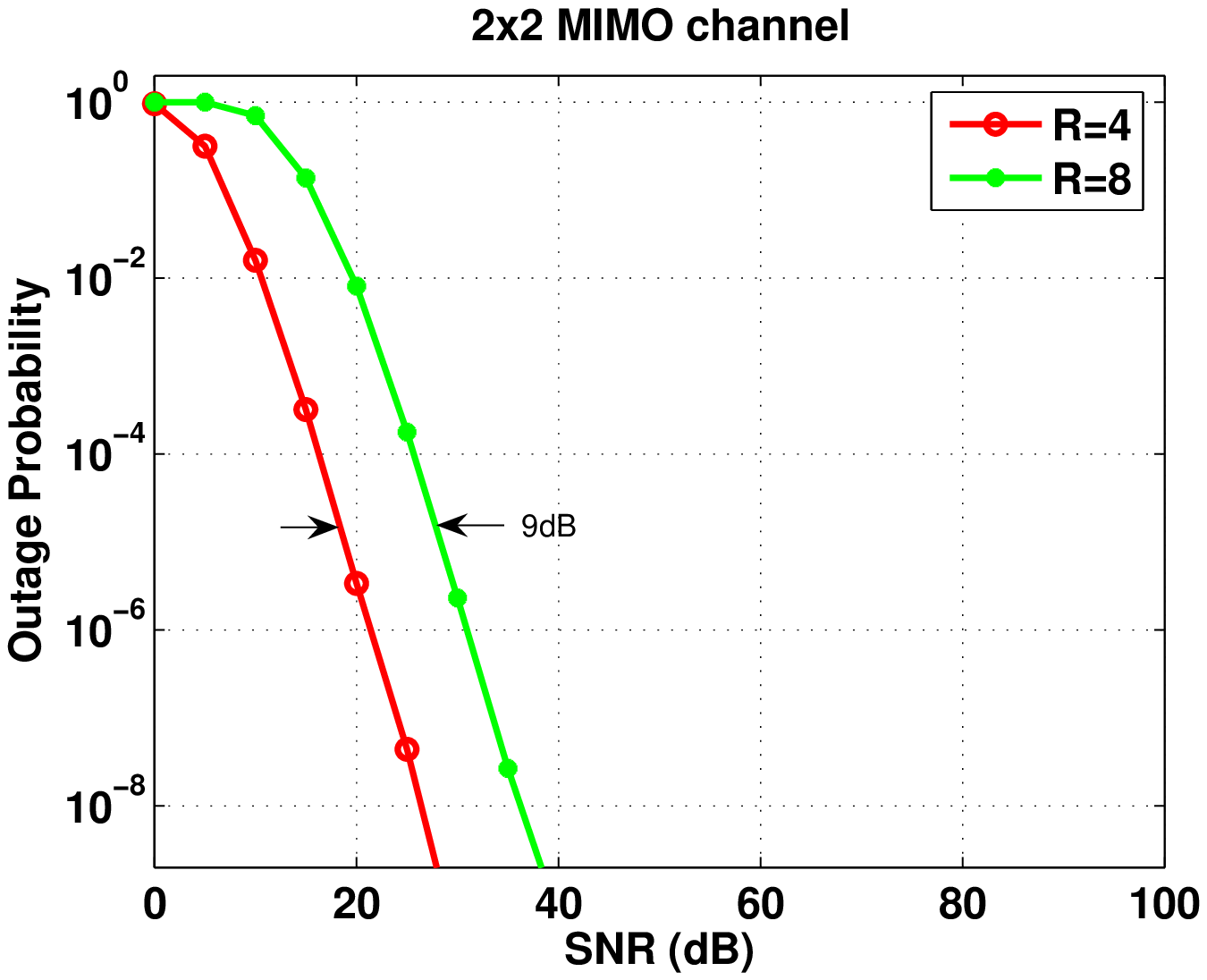}{Outage curves corresponding to $R=4,8$
bpcu, for a $2 \times 2$ MIMO channel.}{5}{}

\putFrag{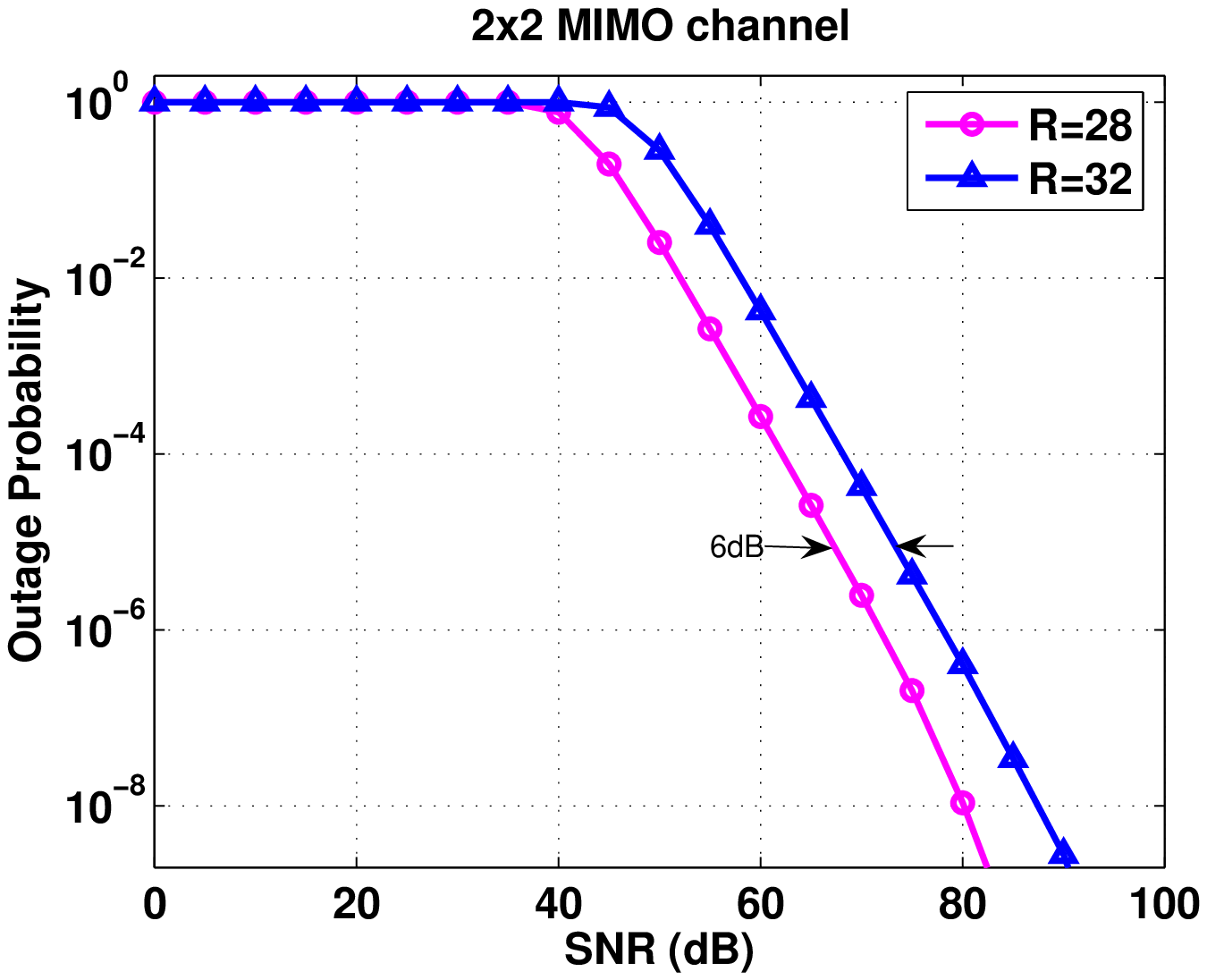}{Outage curves corresponding to $R=28,32$
bpcu, for a $2 \times 2$ MIMO channel.}{5}{}

\putFrag{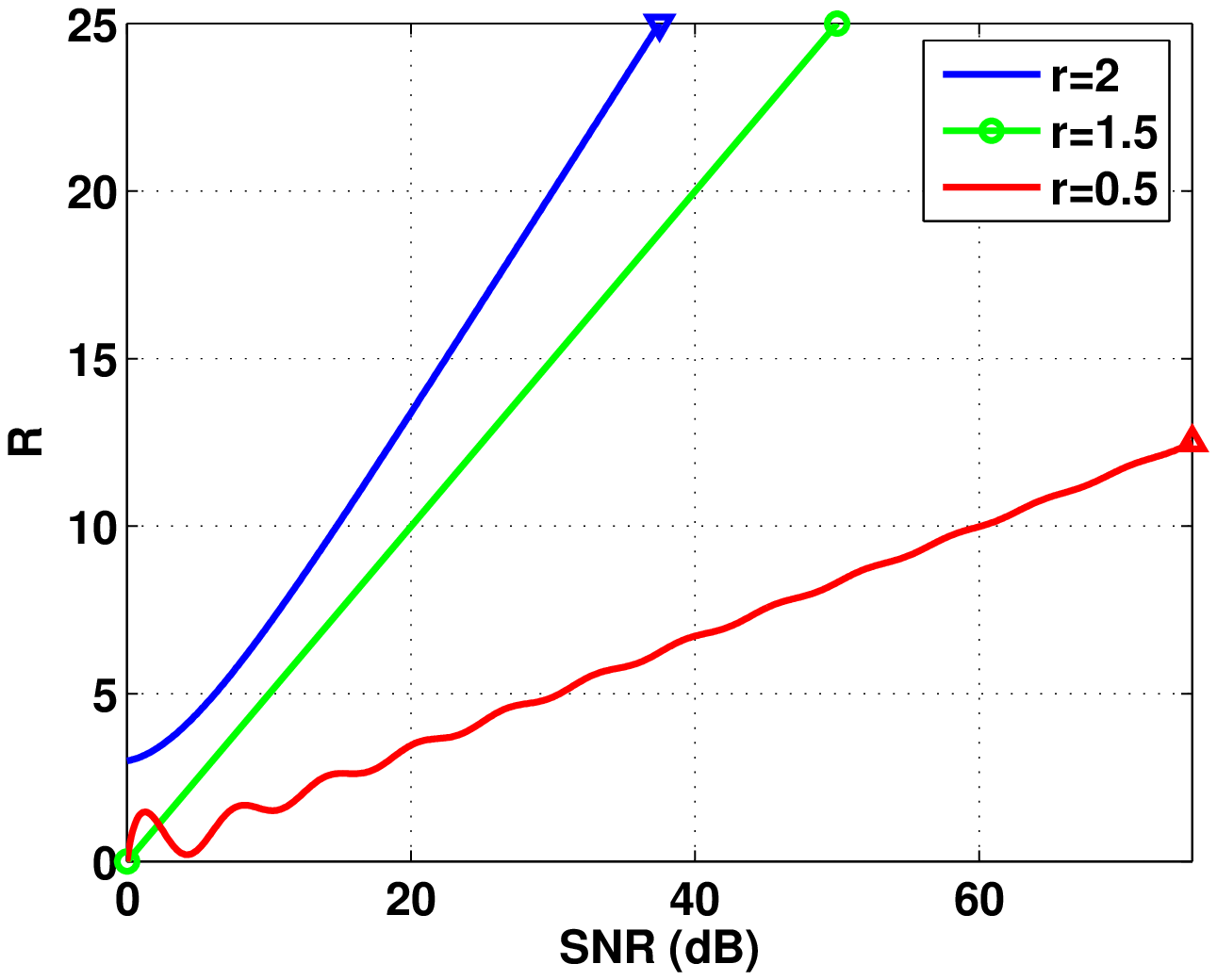}{The notion of multiplexing gain restricts the
scenarios of interest to those in which $R$ \emph{asymptotically}
scales linearly with $\log \rho$, i.e. $R \sim r \log \rho$.}{5}{}

\putFrag{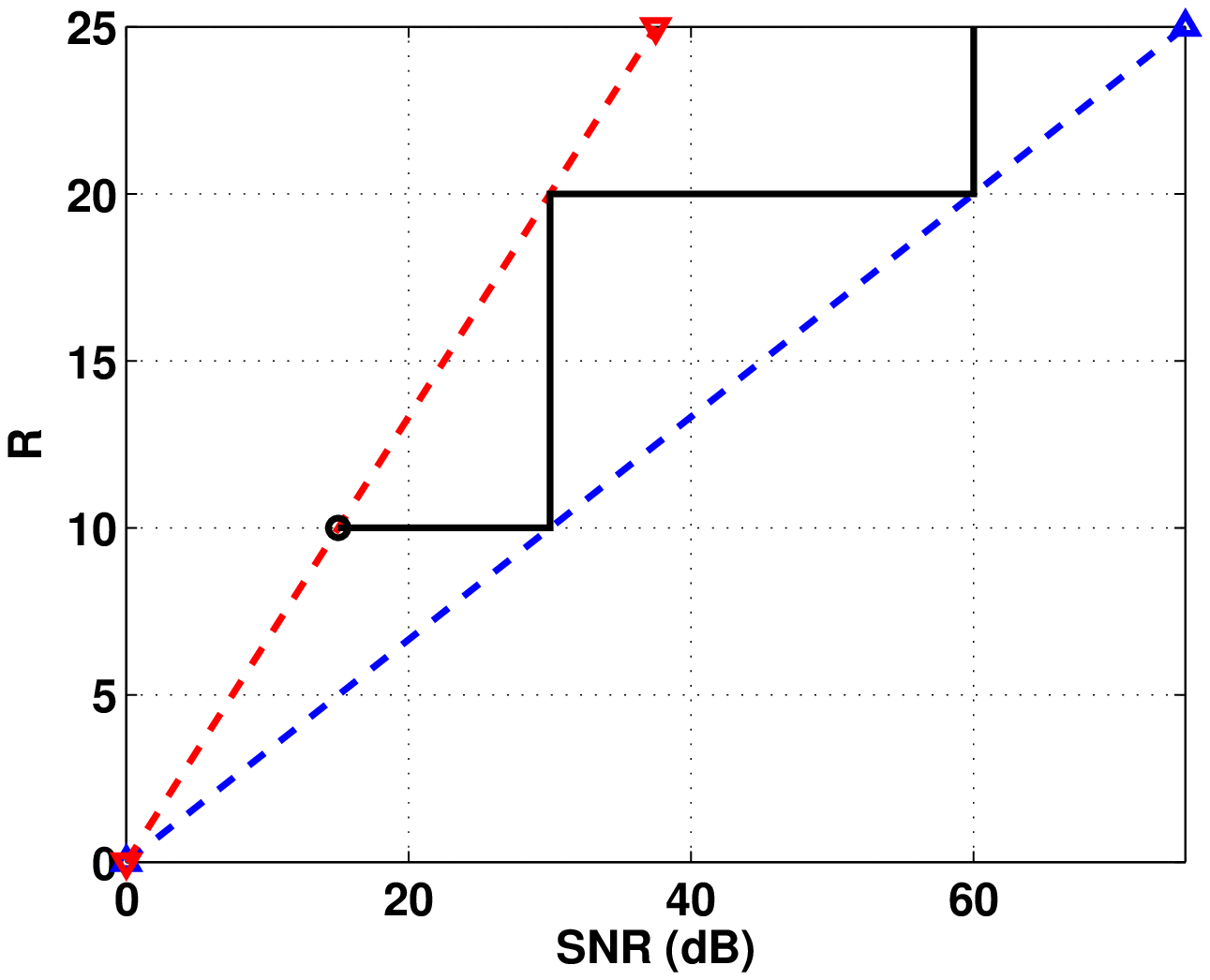}{Relaxing the constraint imposed by the
multiplexing gain notion; A multiplexing gain cannot be defined for
the depicted trajectory, however, since it remains well within an
operating region (i.e. $\mathcal{R}(1)$) TRT analysis can be
applied. }{5}{}

\putFrag{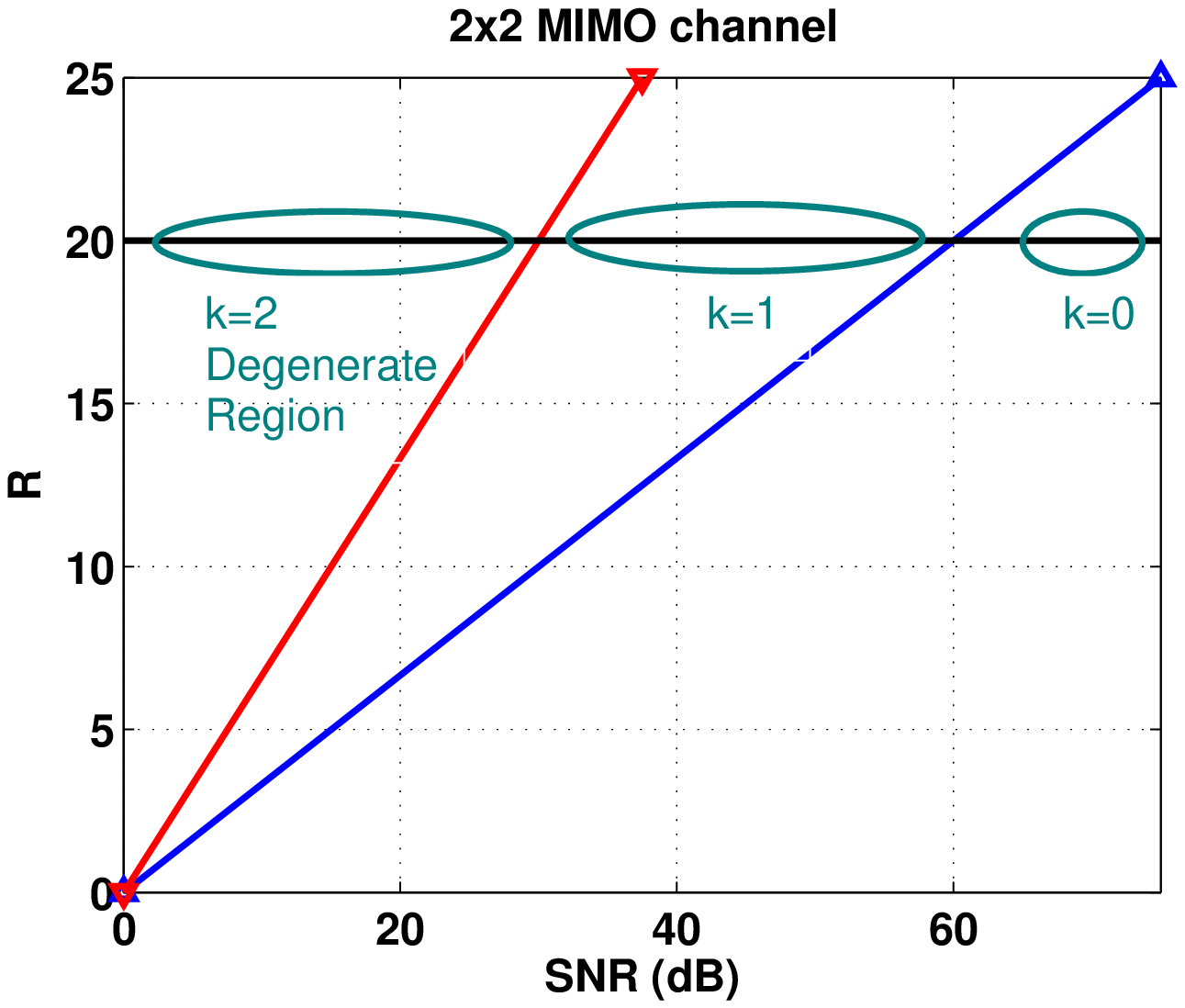}{The constant rate trajectory with $R=20$
bpcu passes through different operating regions in a $2 \times 2$
MIMO system}{5}{}

\putFrag{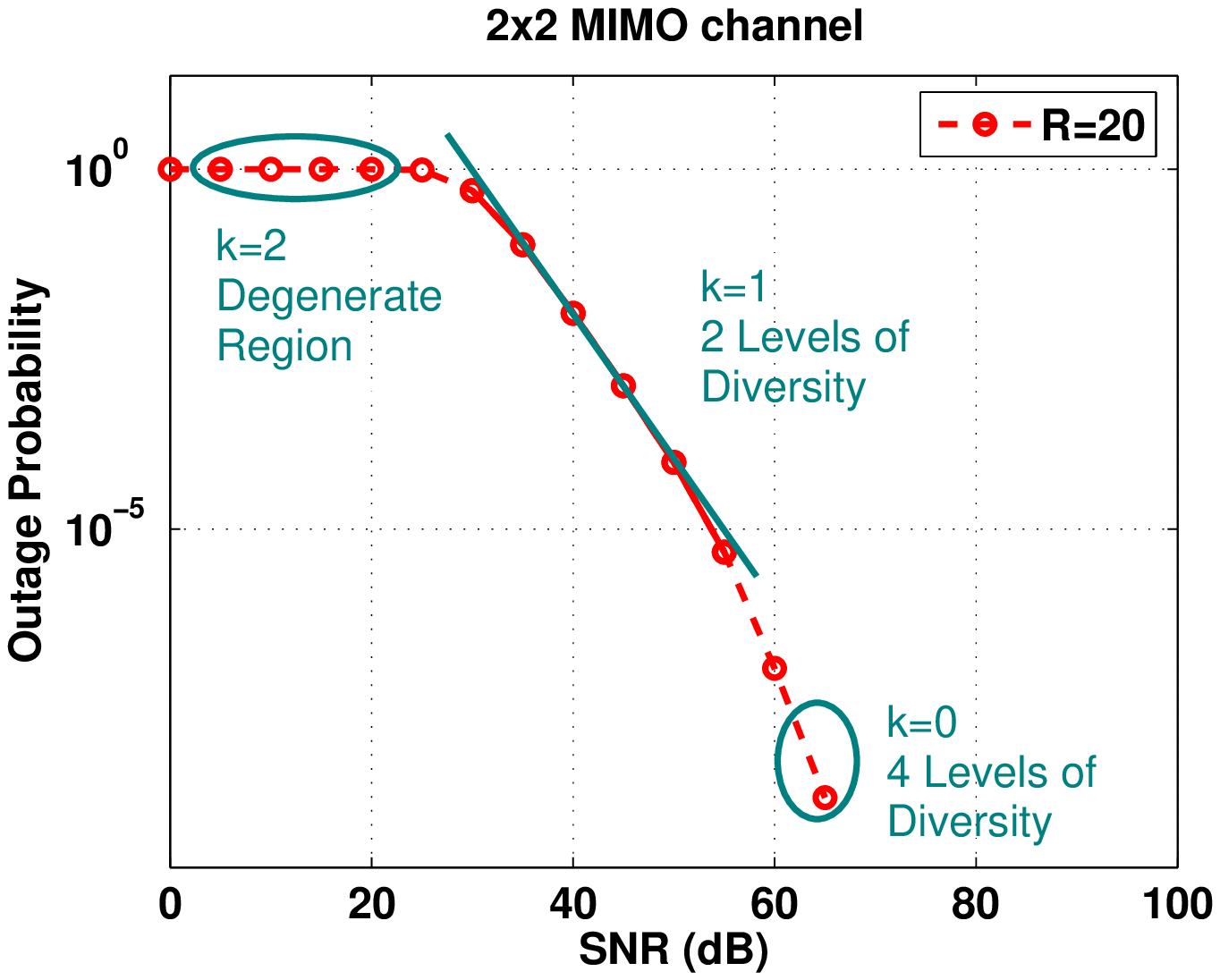}{Outage curves corresponding to $R=20$ bpcu
for a $2 \times 2$ MIMO channel. The solid segment corresponds to
the $\mathcal{R}(1)$ operating region.}{5}{}

\putFrag{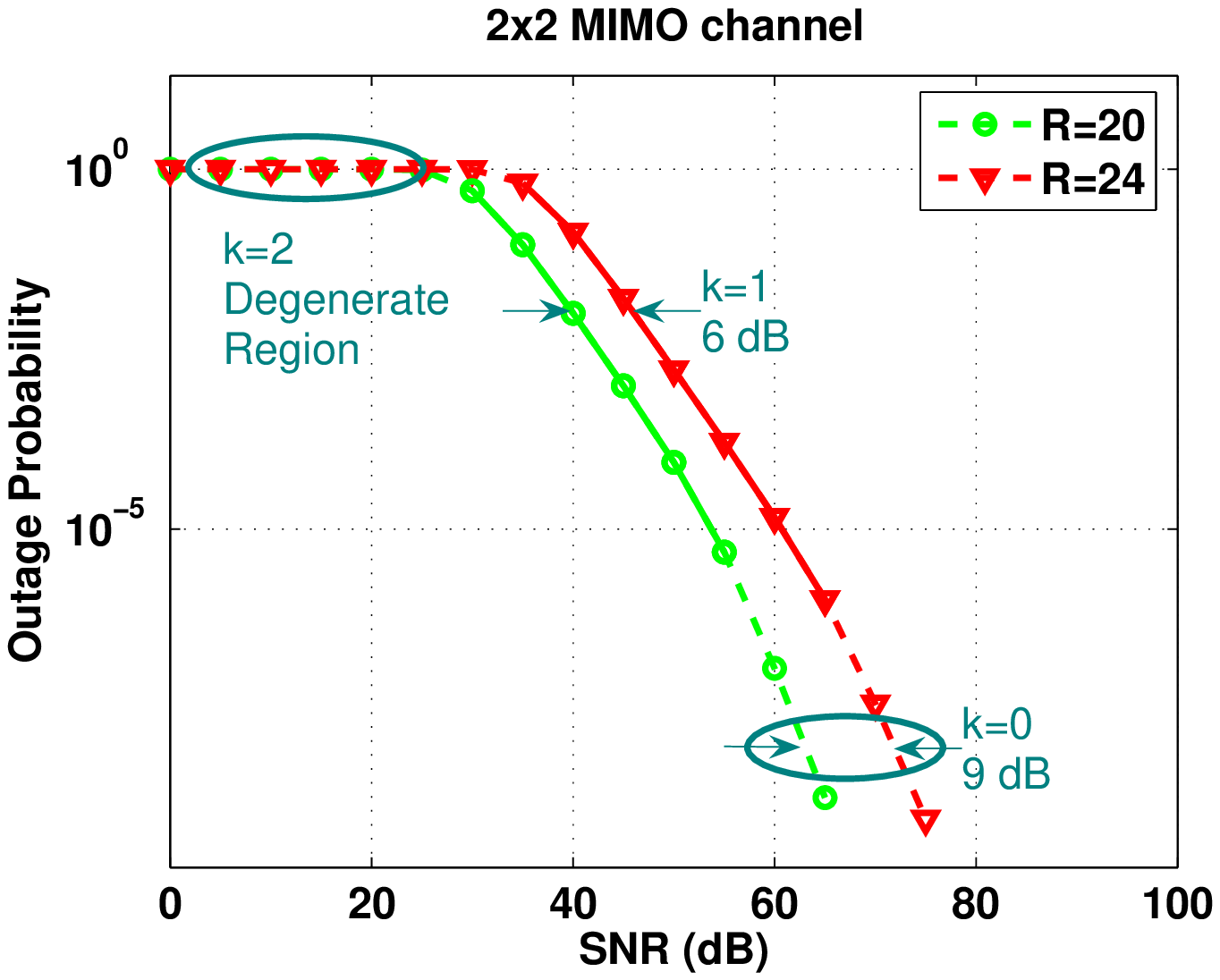}{Outage curves corresponding to $R=20,24$
bpcu for a $2 \times 2$ MIMO channel. The solid segments correspond
to the $\mathcal{R}(1)$ operating region.}{5}{}

\putFrag{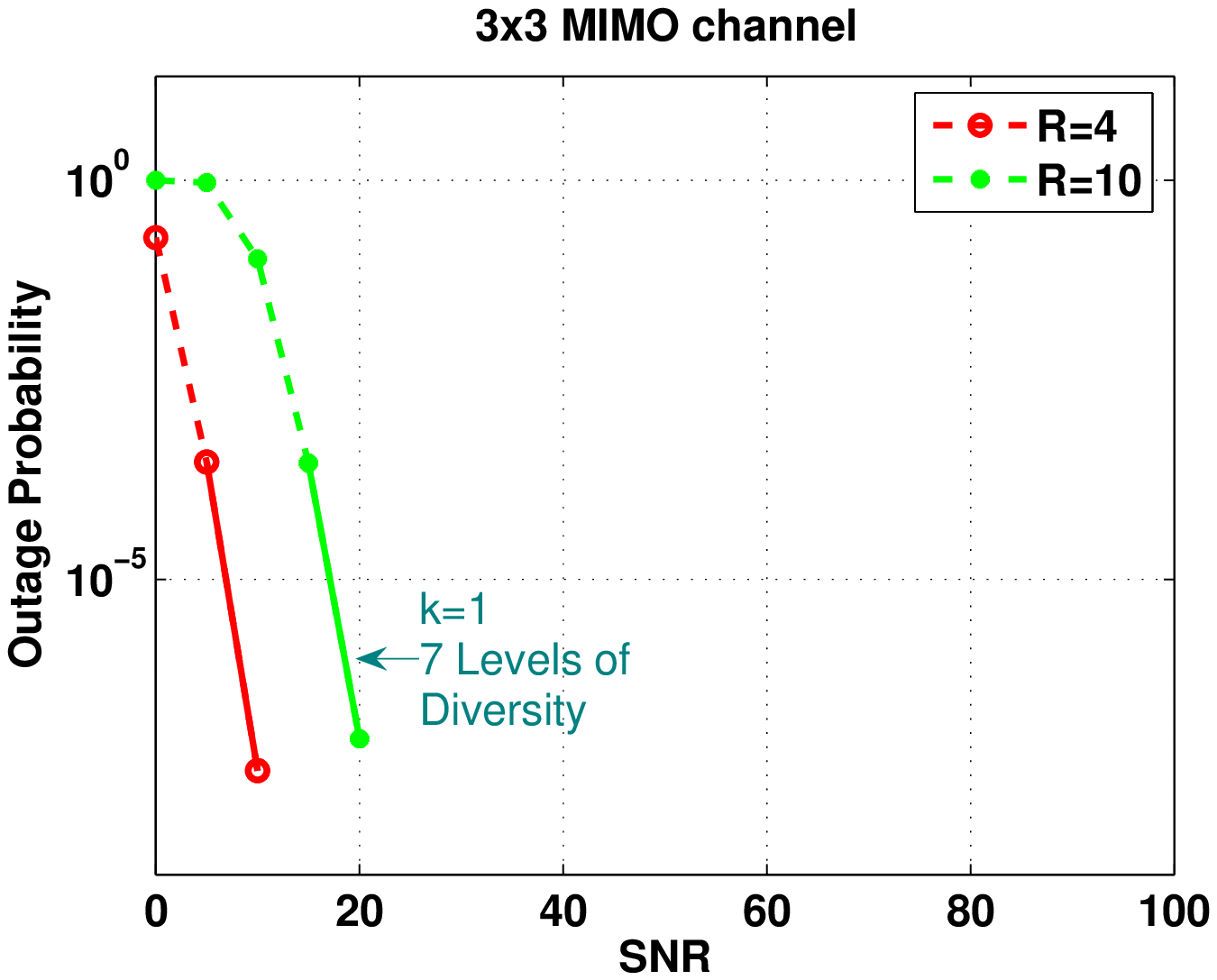}{Outage curves corresponding to $R=4,10$
bpcu for a $3 \times 3$ MIMO channel. The solid segments correspond
to the $\mathcal{R}(1)$ operating region.}{5}{}

\putFrag{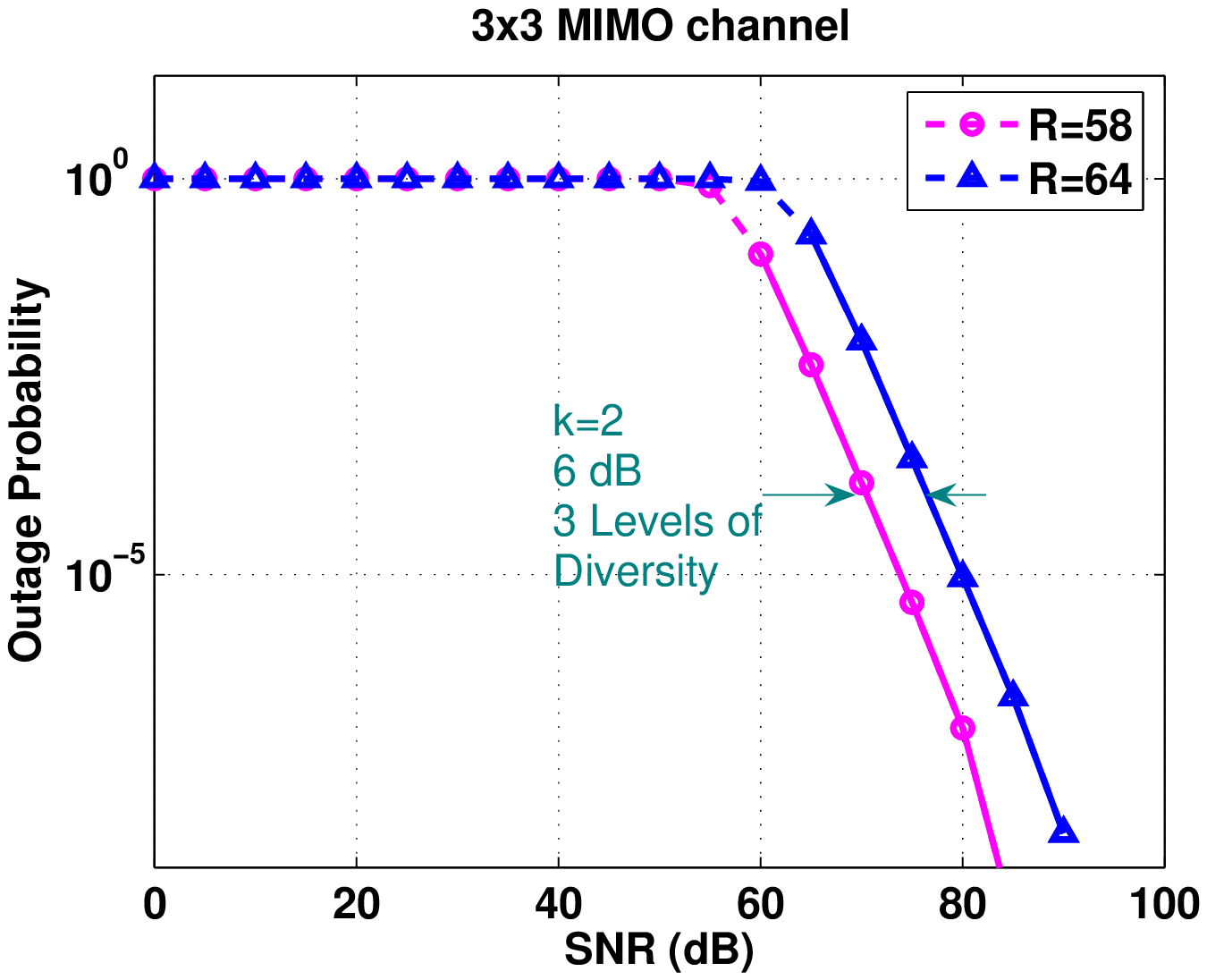}{Outage curves corresponding to $R=58,64$
bpcu for a $3 \times 3$ MIMO channel. The solid segments correspond
to the $\mathcal{R}(2)$ operating region.}{5}{}

\putFrag{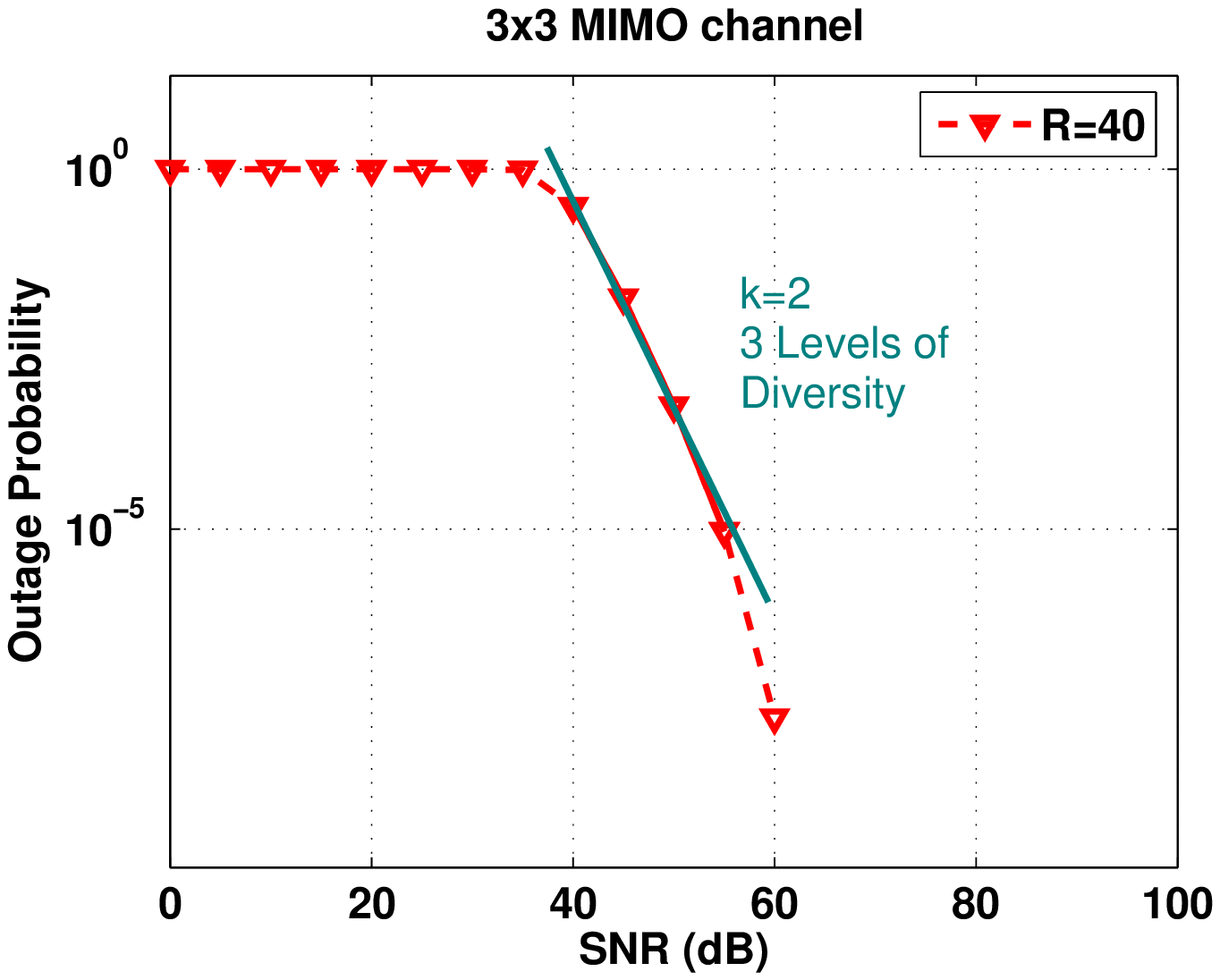}{Outage curves corresponding to $R=40$ bpcu
for a $3 \times 3$ MIMO channel.  The solid segment corresponds to
the $\mathcal{R}(2)$ operating region.}{5}{}

\putFrag{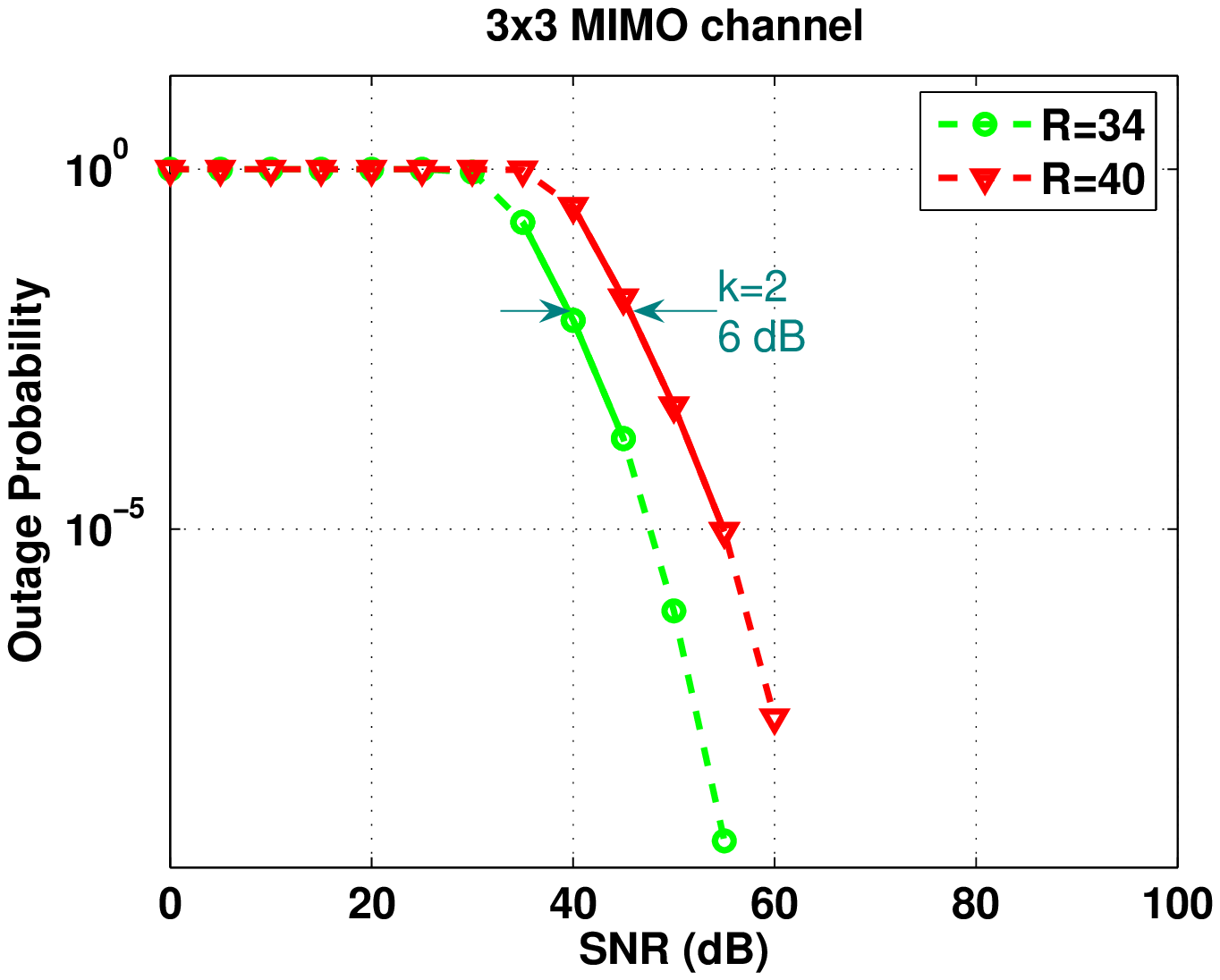}{Outage curves corresponding to $R=34,40$
bpcu for a $3 \times 3$ MIMO channel.  The solid segments correspond
to the $\mathcal{R}(2)$ operating region.}{5}{}

\putFrag{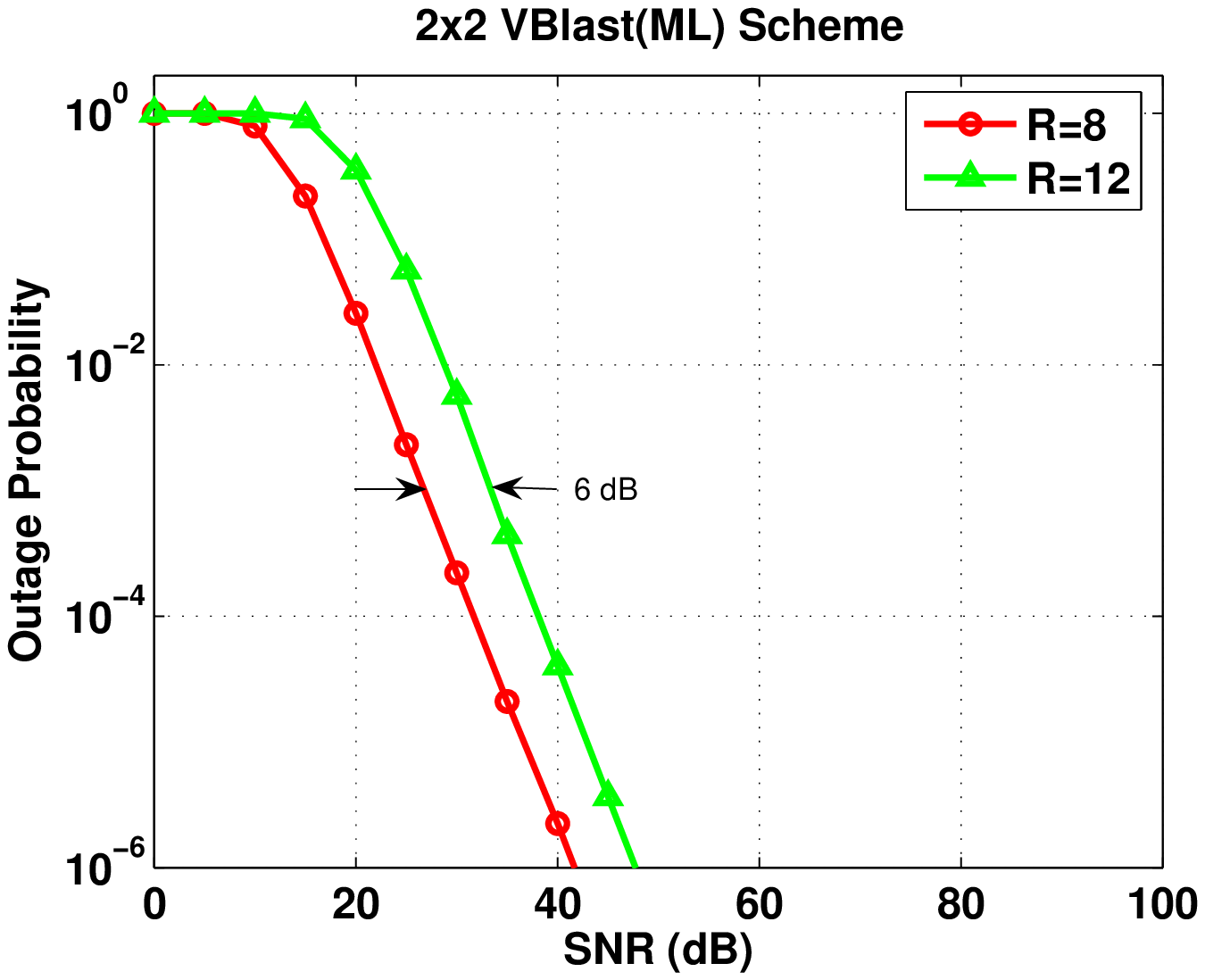}{Outage curves corresponding to
$R=8,12$ bpcu for a $2 \times 2$ V-BLAST scheme.}{5}{}

\putFrag{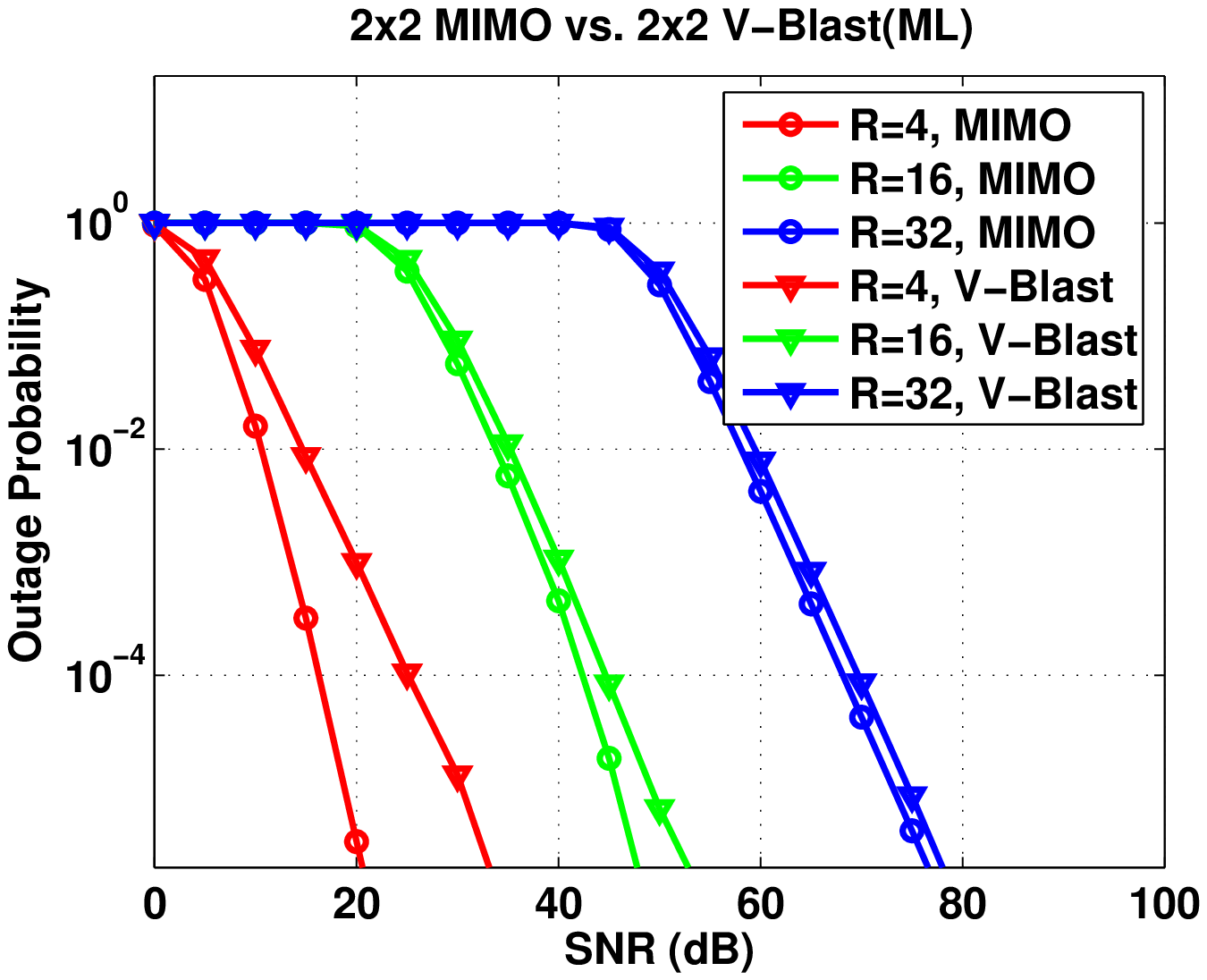}{Comparison of outage curves
corresponding to $R=4,16,32$ bpcu for the $2 \times 2$ MIMO channel
and the V-BLAST scheme.}{5}{}

\putFrag{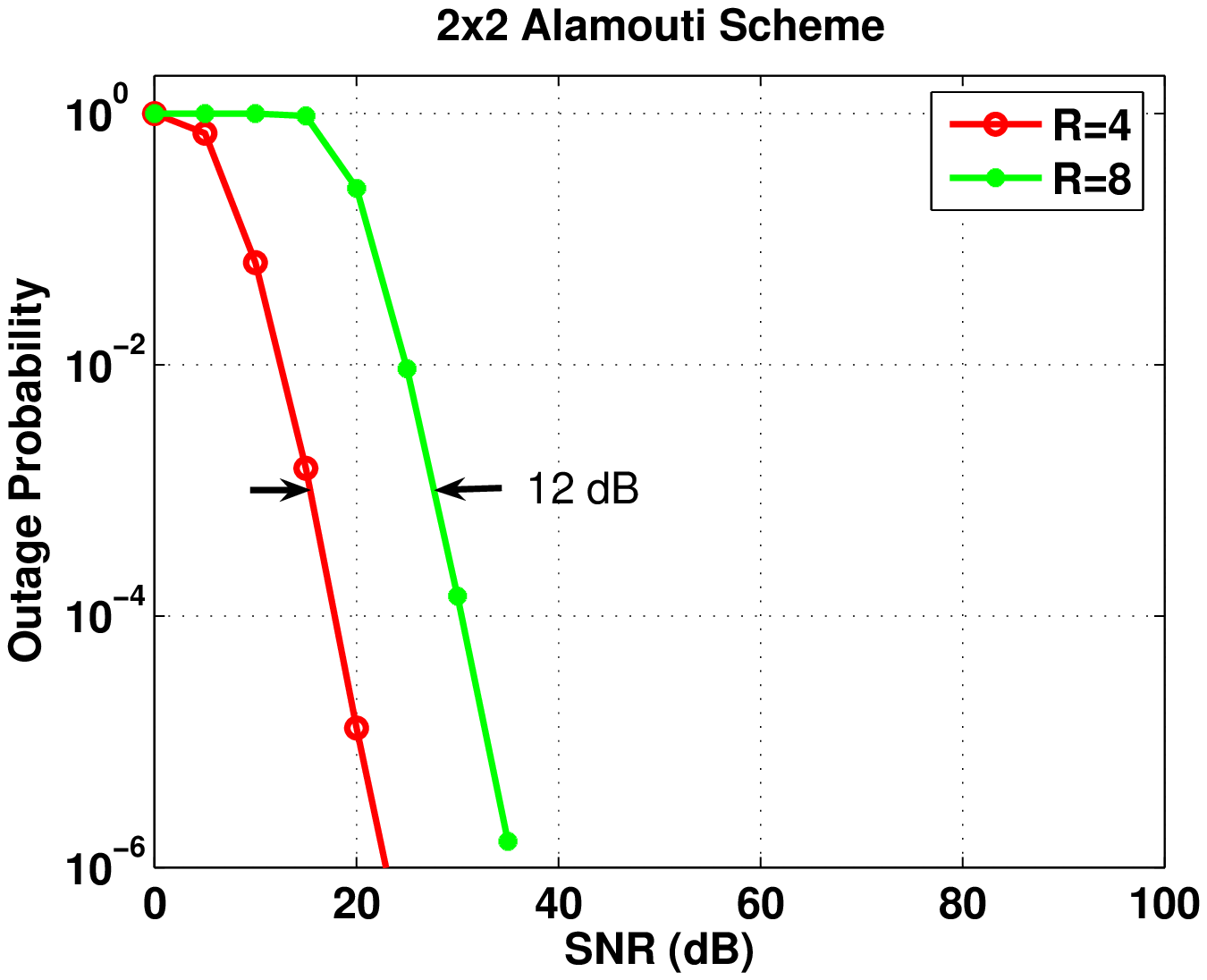}{Outage curves corresponding to
$R=4,8$ bpcu for the $2 \times 2$ Alamouti scheme.}{5}{}

\putFrag{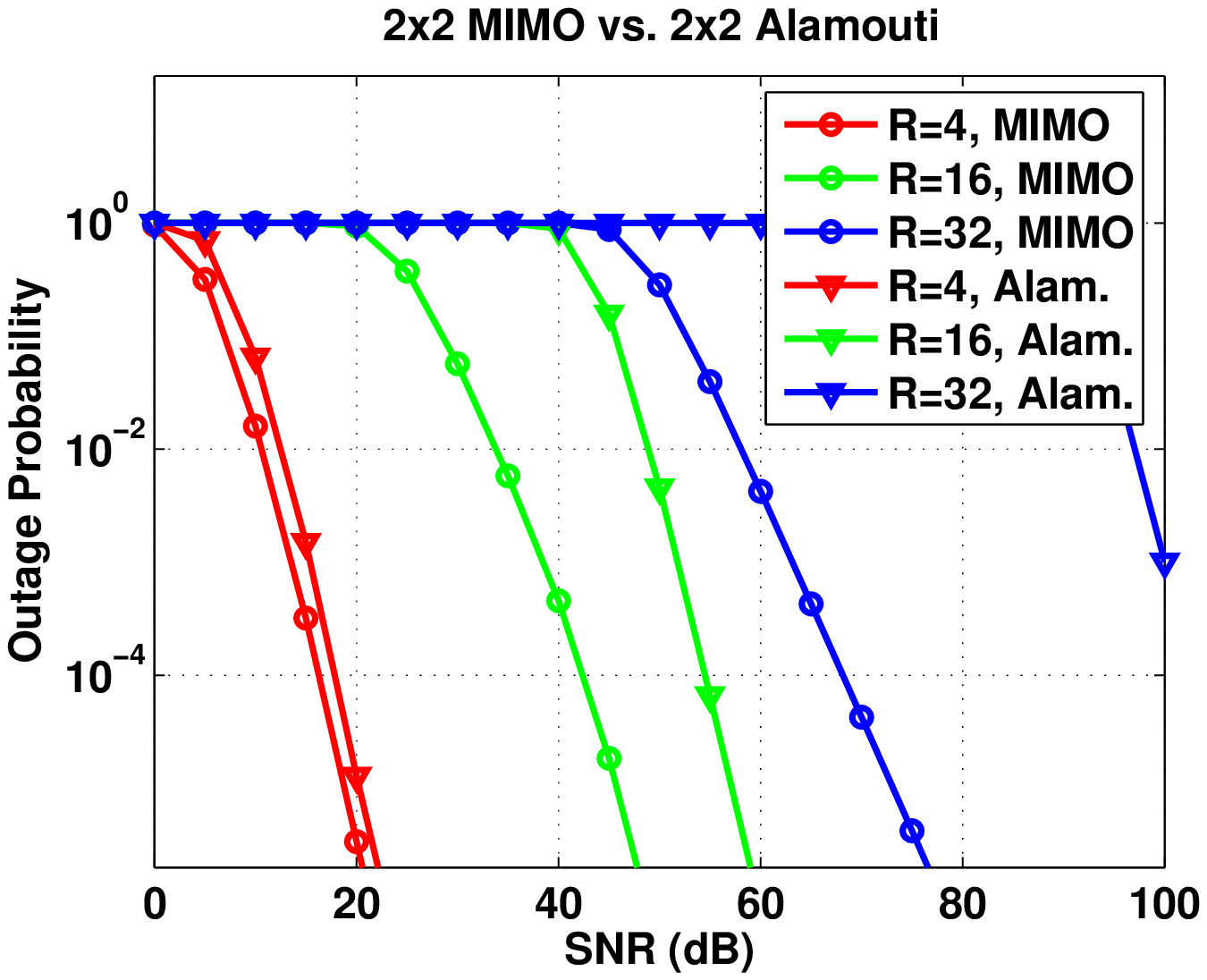}{Comparison of outage curves
corresponding to $R=4,16,32$ bpcu for the $2 \times 2$ MIMO channel
and the Alamouti scheme.}{5}{}


\begin{thebibliography}{1}

\bibitem{ZT02}
L.~Zheng and D.~N.~C. Tse,
\newblock "Diversity and Multiplexing: A Fundamental Tradeoff in Multiple
  Antenna Channels,"
\newblock {\em IEEE Trans. Info. Theory}, 49:1073-1096, May 2003.

\bibitem{FGVW:99}
G.~Foschini, G.~Golden, R.~Valenzuela and P.~Wolniansky, \newblock
"Simplified Processing for High Spectral Efficiency Wireless
Communication Employing Multi-Element Arrayas," \newblock {\em IEEE
Jour. Select. Areas on Comm.}, 17:1841-1852, Nov. 1999.

\bibitem{TVZ:03}
D.~N.~C.~Tse, P.~Viswanath and L.~Zheng,
\newblock "Diversity-multiplexing Tradeoff in Multiple-Access
Channels," \newblock {\em IEEE Trans. Info. Theory}, 50:1859-1874,
Sept. 2004.

\bibitem{TJC:99}
V.~Tarokh, H.~Jafarkhani and A.~R.~Calderbank,
\newblock "Space-Time Block Codes from Orthogonal Designs,"
\newblock {\em IEEE Trans. Info. Theory}, 45:1456-1467, July 1999.

\bibitem{A:98}
S.~Alamouti, \newblock "A Simple Transmitter Diversity Scheme for
Wireless Communications," \newblock {\em IEEE Jour. Select. Areas on
Comm.}, 16:1451-1458, Oct. 1998.

\bibitem{ECD:04}
H.~El~Gamal, G.~Caire and M.~O.~Damen,
\newblock "The MIMO ARQ Channel: Diversity-Multiplexing-Delay
Tradeoff," \newblock {\em Submitted to the IEEE Trans. Info.
Theorey}.

\end{thebibliography}
\end{document}